\documentclass[aps,pre,twocolumn,showpacs]{revtex4}
\usepackage{float}
\usepackage{graphicx,amssymb,amsmath}
\usepackage{natbib}
\usepackage{bm}
\begin{document}
\title{Short-time inertial response of viscoelastic fluids measured with Brownian motion and with active probes}
\author{M.\ Atakhorrami$^{1,2}$}
\author{D.\ Mizuno$^{1,3}$}
\author{G.H.\ Koenderink$^{1,4}$}
\author{T.B.\ Liverpool$^{5,6,7}$}
\author{F.C.\ MacKintosh$^{1,6}$}
\author{C.F.\ Schmidt$^{1,8}$}

\affiliation{$^{1}$Division of Physics and Astronomy, Vrije
Universiteit 1081HV Amsterdam, The Netherlands}

\affiliation{$^{2}$Philips Research, 5656AE Eindhoven, The Netherlands}

\affiliation{$^{3}$Organization for the Promotion of Advanced Research,
Kyushu University, 812-0054 Fukuoka, Japan}

\affiliation{$^{4}$FOM Institute AMOLF, 1098SJ Amsterdam, The Netherlands}

\affiliation{$^{5}$Department of Applied Mathematics, University
of Leeds, Leeds, LS2 9JT, United Kingdom}

\affiliation{$^{6}$Isaac Newton Institute for Mathematical
Sciences, University of Cambridge, Cambridge, CB3 0EH, UK}

\affiliation{$^{7}$Department of Mathematics, University of Bristol, Bristol BS8 1TW, UK}

\affiliation{$^{8}$Fakult\"at f\"ur Physik, Georg-August-Universit\"at, 37077 G\"ottingen, Germany}

\date{\today}

\begin{abstract}
We have directly observed short-time stress propagation in viscoelastic fluids using
two optically trapped particles and a fast interferometric
particle-tracking technique. We have done this both by recording
correlations in the thermal motion of the particles and
by measuring the response of one particle to the actively
oscillated second particle. Both methods detect the vortex-like
flow patterns associated with stress propagation in fluids.
This inertial vortex flow propagates diffusively for
simple liquids, while for viscoelastic solutions the pattern
spreads super-diffusively, dependent on the shear modulus of the
medium.
\end{abstract}

 \maketitle

\maketitle

\section{Introduction\label{intro}}

Motion in simple liquids at small scales is usually characterized by low Reynolds numbers,
in which the response of a liquid to a force applied at one point is Stokes-like---decaying with distance $r$ as $1/r$ away from the origin of the disturbance \cite{LandauFluids,brenner,Guyon}. Here, fluid inertia can be neglected, and the force is effectively
felt instantaneously everywhere within the medium. In practice, this is a good approximation, for instance, in water at the colloidal scale up to micrometers on times scales larger than a microsecond. At short times or high frequencies, however, fluid inertia limits the range of stress propagation. Any instantaneous disturbance must be confined to a small region for short times. If the medium is also incompressible, then this naturally gives rise to vorticity and  backflow. In simple liquids, stress then propagates through the
diffusive spreading of this vortex. Although this basic physical
picture has been known theoretically for simple liquids since the
work of Oseen in 1927~\cite{Oseen} and has been shown in computer simulations since the
1960s~\cite{Alder}, experimental observation of these effects have been largely indirect, for instance, in the form of short-time corrections to Brownian motion \cite{exper_tails}. Direct experimental observation has only recently been possible because of the high temporal and spatial resolution required \cite{maryam}.

The finite time it takes for vorticity to propagate leads to
persistence of fluid motion that manifests itself in
algebraic decay of the auto-correlation function of the velocity of
either a fluid element or a particle embedded in the fluid. This decay is slower than the naive expectation of exponentially-decaying correlations for a massive particle experiencing viscous drag. Thus, the effect is known as the \emph{long time tail} effect~\cite{exper_tails,theor_tails,Lukic}, which characterizes the transition from
ballistic to Brownian motion of particles in simple liquids. This effect has been shown to be present even at the atomic level, \emph{e.g.}, from neutron scattering experiments on
liquid sodium~\cite{Morkel}.

We have shown that the inter-particle correlations and response
functions of two particles can be used to directly resolve the flow pattern and dynamics of vortex propagation \cite{maryam,liverpool}.
This was done by measuring the correlated thermal motion
of two optically trapped spherical particles using an
interferometric technique ~\cite{detection Gittes,setup maryam}
with high temporal (up to 100~kHz) and spatial (sub-nanometer)
resolution in both viscous and viscoelastic fluids. We were able
to observe, for instance, the anti-correlation in the inter-particle fluctuations
of thermal motion that is characteristic of
the vortex propagation. The method is related to passive
two-particle microrheology ~\cite{crocker, Starrs, meiners,
bartlett, Gardel, Koenderink actin} which can be used to measure
shear elastic moduli of viscoelastic materials.

The inter-particle response functions $\alpha (\omega)$, with
real ($\alpha'$) and imaginary ($\alpha''$) parts defined by
$\alpha(\omega)=\alpha'(\omega)+i\alpha''(\omega)$ are obtained
for motion parallel ($\alpha_{||}$) and perpendicular
($\alpha_{\bot}$) to the centerline connecting the two particles.
In the passive approach, we directly measure the imaginary part of
the response function from the thermal position fluctuations of
the two particles via the fluctuation-dissipation theorem (FDT).
The real part of the response function is then obtained from a
Kramers-Kronig integral~\cite{MR97}.

Here, in order to directly measure both real (in-phase) and imaginary (out-of-phase) parts of
the response function, we have developed an active method
~\cite{Hough,Mizuno active}, in which one optical trap
drives oscillatory motion of one particle, while the response of a second particle is measured at separation $r$. We present and compare detailed experimental results of both passive and
active approaches. We also present a theoretical derivation of
the predicted response functions and corresponding
algebraic decay of the velocity autocorrelation functions for
viscoelastic fluids.

The outline of the paper is as follows. In
section II, we present the theoretical analysis. In section III,
we present the materials and methods of sample preparation, as
well as the experimental techniques for the passive and active
measurements of the response functions. In section IV we describe
our methods of data analysis used for the results presented in
section V. In the results section V, we first compare the data for
simple liquids with the dynamic Oseen tensor, which demonstrates
the diffusive propagation of the vortex flow. We then present our
results and comparison with theory for viscoelastic solutions,
including the evidence for superdiffusive stress propagation.
Finally, we conclude with a discussion (section VI), also
mentioning implications of our results for microrheology in
general.

\section{THEORY AND BACKGROUND \label{theo}}

Newtonian liquids are described by the Navier-Stokes equation,
which is non-linear. The non-linearity, however, can usually be
neglected either for small distances or for low
velocities~\cite{LandauFluids,brenner}. This is the so-called low
Reynolds number regime, since the relative importance of
non-linearities is characterized by the Reynolds number $\mbox{Re}
= \frac{UL \rho}{\eta}$, where $U$, $L$, $\rho$, and $\eta$ are,
respectively, the characteristic velocity and length scales, the
density, and the viscosity. For steady flow, this regime can also
be thought of as the non-inertial regime, in which stress
propagates instantaneously and, for instance, the velocity
response at a distance $r$ from a point force varies as
$1/r$~\cite{LandauFluids,brenner,Guyon}. Such \emph{Stokes} flow
accurately describes the motion of micron-size objects in water on
time scales longer than a few microseconds.

Even at low Reynolds number, however, there are remaining
consequences of fluid inertia for non-stationary
flows~\cite{Guyon}. This {\em unsteady Stokes} approximation is
described by the linearized Navier-Stokes equation:
\begin{equation}
\rho\frac{\partial}{\partial t}\vec v=\eta\nabla^2\vec
v-\vec\nabla P+\vec f, \label{eq:linNS}
\end{equation}
where $\vec v$ is the velocity field, $P$ is the pressure that
enforces the incompressibility of the liquid and $\vec{f}$ is the
force density {\em applied} to the fluid. By taking the curl of
this equation we observe that the vorticity
$\vec\Omega=\vec\nabla\times\vec v$ satisfies the diffusion
equation with diffusion constant $\nu=\eta/\rho$.

As noted above, the short-time response of a liquid to a point
force generates a vortex. The propagation of stress away from the
point disturbance is diffusive: after a time $t$, this vortex
expands away from the point force to a size of order
$\delta\sim\sqrt{\eta t/\rho}$. In the wake of this moving vortex
is the usual Stokes flow that corresponds to a $1/r$ dependence of
the velocity field. For an oscillatory disturbance at frequency
$\omega$, the propagation of vorticity defines a \emph{penetration
depth} $\delta\sim\sqrt{\eta/(\omega\rho)}$~\cite{LandauFluids}.
Stress effectively propagates instantaneously on length scales
shorter than this.

This picture generalizes to the case of homogenous
\emph{viscoelastic} media characterized by an isotropic,
time-dependent shear modulus $G(t)$~\cite{bird}, although the propagation
of stress generally becomes super-diffusive \cite{liverpool}. We further assume
that the medium is incompressible, which is a particularly good
approximation for polymer solutions such as those considered here,
at least at high frequencies~\cite{Brochard,MR97,2fluid}. The
deformation of the medium is characterized by a {\em local}
displacement field $\vec{u}(\vec{r},t)$, and the viscoelastic
analogue of the Navier-Stokes equation (\ref{eq:linNS}) is:
\begin{eqnarray}
\rho\frac{\partial^2}{\partial t^2}\vec u(\vec{r},t) &=&
\vec{\nabla} \cdot \tensor{\bm\sigma} (\vec{r},t) -
\vec{\nabla} P + \vec{f}(\vec{r},t) \, , \label{unsteady_ve}
\end{eqnarray}
where
\begin{eqnarray}
\tensor{\bm \sigma}(\vec{r},t) &=& 2 \int^t_{-\infty} dt' G(t-t')
\tensor{\bm \gamma}(\vec{r},t') \label{constitutive}
\end{eqnarray}
is the local stress tensor and
\begin{equation}
\tensor{\bm \gamma} = \frac{1}{2}\left[ \vec{\nabla}\vec{u} +
\left(
    \vec{\nabla}\vec{u}\right)^\dagger\right]
\end{equation}
is the local deformation tensor. Incompressibility corresponds to
the constraint ${\vec \nabla \cdot \vec u}=0$

Equations (\ref{unsteady_ve},\ref{constitutive}) can be simplified
by a decomposition of the force density and deformation into
Fourier components.  Taking spatio-temporal Fourier Transforms
defined as
\begin{equation} \vec u(\vec{k},\omega) = \int d^3 r
\int_{-\infty}^\infty dt\, \, e^{i (\omega t - \vec k \cdot \vec
r)}\vec u(\vec{r},t),
\end{equation}
and defining the complex modulus
\begin{equation}
G^\star(\omega) \equiv G'(\omega) + i G''(\omega) = \int_0^\infty
dt e^{i \omega t} G(t),
\end{equation}
we can eliminate the pressure by imposing incompressibility in
Eqs.\ (\ref{unsteady_ve},\ref{constitutive}). This leads to
\begin{equation}
\vec u(\vec k,\omega)=\left(\frac{{\bf 1} -\hat k\hat
k}{G^\star(\omega) k^2-\rho\omega^2}\right)\cdot \vec f(\vec
k,\omega) \, ,\label{G-Oseen}
\end{equation}
where $\hat k = \vec k / |k|$.  We invert this Fourier transform
to obtain the displacement response function due to a point force
applied at the origin.

\subsection{Response functions}
For a point force $\vec f$ at the origin, the linear response of
the medium at $\vec r$ is given by a tensor $\alpha_{ij}$, where
$u_i (\vec r,\omega) = \alpha_{ij}(\vec r , \omega) f_j(\vec 0,
\omega) $. Here, $\alpha_{ij}= \alpha'_{ij}+i\alpha''_{ij}$ is, in
general, complex. Given our assumptions of rotational and
translational symmetry, there are only two distinct components of
the response. These are (1) a \emph{parallel} response that is
given by a displacement field $\vec u$ parallel to both $\vec f$
and $\vec r$, and (2) a \emph{perpendicular} response given by
$\vec u$ parallel to $\vec f$ and perpendicular to $\vec r$.
The parallel response function
$\alpha_\parallel$, for instance, is obtained from the inverse
Fourier transform of Eq.\ (\ref{G-Oseen}) \cite{liverpool}.

The response functions for general $G^\star(\omega)$ are given by
\begin{equation} \alpha_\parallel\left(r,\omega\right)=
\alpha_\parallel' + i \, \alpha_\parallel''= \frac{1}{4\pi
G^\star(\omega) r} \chi_\parallel
\left(r\sqrt{\kappa}\right),\label{alpha-par}
\end{equation}
and
\begin{equation}
\alpha_\perp\left(r,\omega\right)= \alpha_\perp'+ i \,
\alpha_\perp'' = \frac{1}{8\pi G^\star(\omega) r} \chi_\perp
\left(r\sqrt{\kappa}\right),\label{alpha-perp}
\end{equation}
where $\kappa={\rho\omega^2/G^\star(\omega)}$ is complex  and
\begin{equation} \chi_\parallel\left(x\right)= \frac{2}{x^2}
\left[\left(1-ix\right)e^{ix}-1\right], \label{chi_parr}
\end{equation}
and
\begin{equation}
\chi_\perp\left(x\right)= \frac{2}{x^2}
\left[1+\left(x^2-1+ix\right)e^{ix}\right]. \label{chi_perp}
\end{equation}
The magnitude of $\kappa$ defines the inverse (viscoelastic)
penetration depth $\delta$.

\subsubsection{Simple liquids}
For a simple liquid, $G^\star(\omega)= - i\,\omega \eta $ and
$\kappa= i \rho \omega / \eta $. The velocity response of the
liquid is then characterized by
\begin{equation}
-i\omega\alpha_\parallel\left(r,\omega\right)=
\omega\alpha_\parallel'' - i\, \omega \alpha_\parallel'=
\frac{1}{4\pi \eta r} \tilde\chi_\parallel
\left(r\sqrt{\frac{\rho\omega}{2\eta}}\right),\label{velocity-par}
\end{equation}
and
\begin{equation}
-i\omega\alpha_\perp\left(r,\omega\right)= \omega\alpha_\perp''- i
\, \omega\alpha_\perp' = \frac{1}{8\pi \eta r} \tilde\chi_\perp
\left(r\sqrt{\frac{\rho\omega}{2\eta}}\right),\label{velocity-perp}
\end{equation}
where
\begin{eqnarray}
&&\tilde\chi_\parallel'\left(x\right)=
\frac{\left[\left(1+x\right)\sin x-x\cos x\right]e^{-x}}{x^2},
\label{chi_parr1}\\
&&\tilde\chi_\parallel''\left(x\right)=
\frac{1-\left[\left(1+x\right)\cos x+x\sin x\right]e^{-x}}{x^2},
\label{chi_parr2}\\
&&\tilde\chi_\perp'\left(x\right)=
\frac{\left[\left(x+2x^2\right)\cos x-\left(1+x\right)\sin
x\right]e^{-x}}{x^2}, \label{chi_perp1}\\
&&\tilde\chi_\perp''\left(x\right)=
\frac{\left[\left(x+2x^2\right)\sin x+\left(1+x\right)\cos
x\right]e^{-x}-1}{x^2}.\nonumber \\ \label{chi_perp2}
\end{eqnarray}
Thus, for instance, the in-phase and out-of-phase velocity
response in the parallel direction are given by
$\omega\alpha''=\frac{1}{4\pi\eta r}\tilde\chi'_\parallel$ and
$-\omega\alpha'=\frac{1}{4\pi\eta r}\tilde\chi''_\parallel$.

\begin{figure}[ht]\centering
\includegraphics[width=8cm]{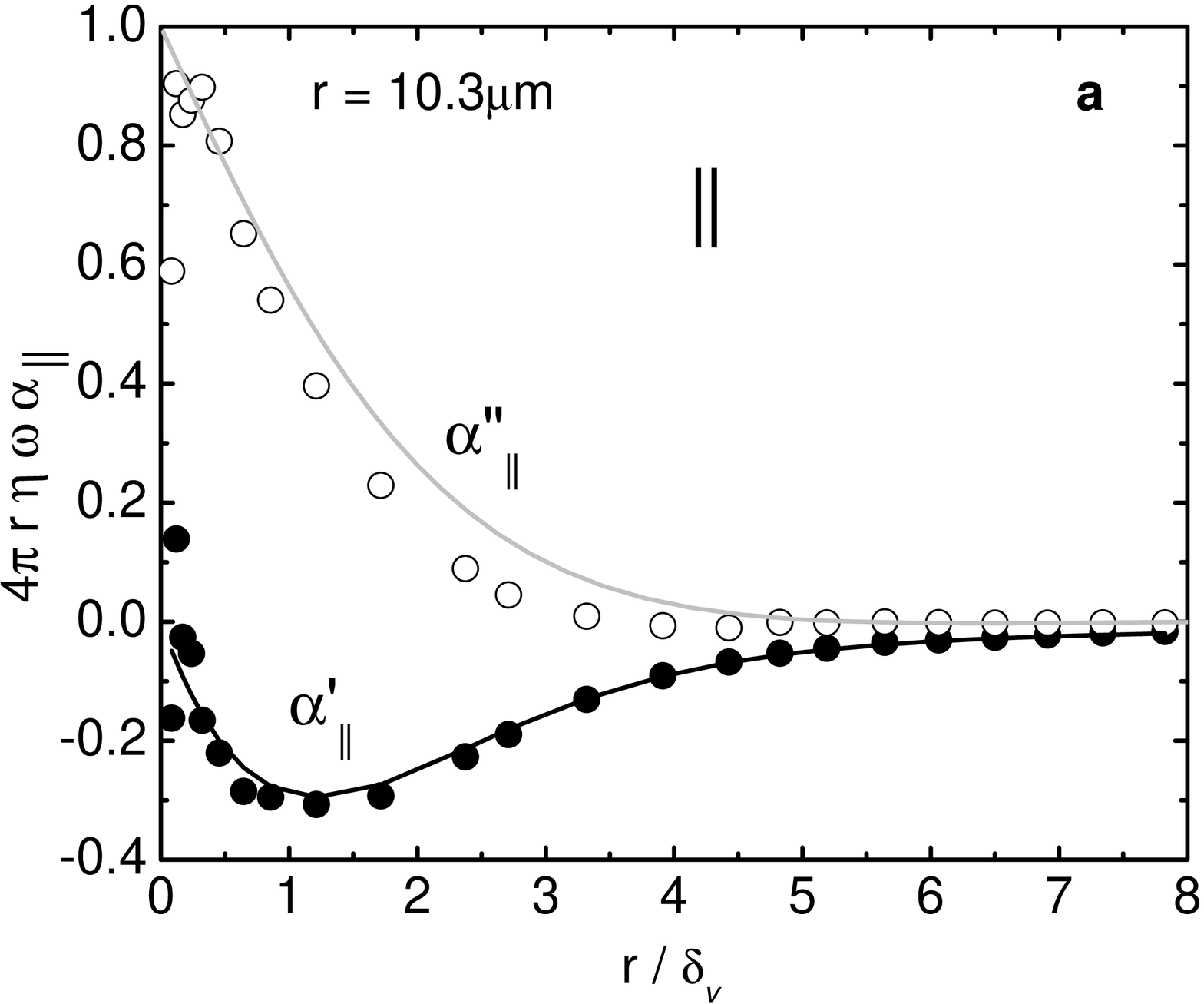}
\includegraphics[width=8cm]{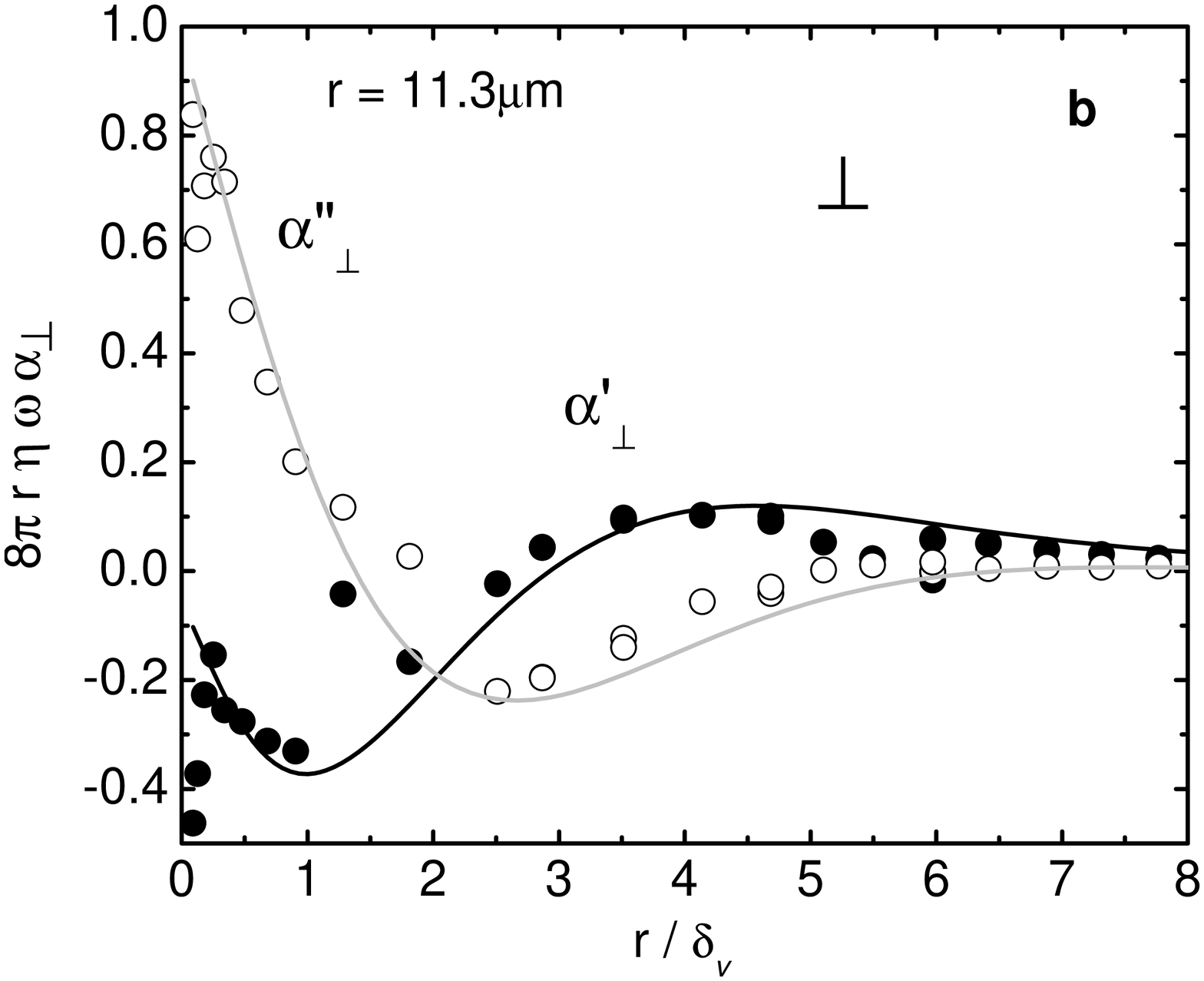}
\caption{Comparison of theoretical and experimentally measured response functions for a simple liquid. The predictions of the normalized velocity field from the dynamic Oseen tensor in Eqs.\ (\ref{velocity-par},\ref{velocity-perp}) are shown as black and gray lines.
Normalized complex inter-particle response functions
between two probe particles (silica beads, $R$ = 0.58~$\mu$m)
measured with the active microrheology method in water: (a) $4 \pi r \eta \omega
\alpha_{||}$ in the parallel direction and (b) $8 \pi r \eta
\omega \alpha_{\bot}$ in the perpendicular direction, plotted
versus the ratio of the separation distance $r$ (fixed for a given
bead pair, $r$ = 11.3~$\mu$m for parallel, and $r$ = 10.3~$\mu$m
for perpendicular) to the frequency-dependent viscous penetration
depth $\delta_{v}$. Both real (filled symbols) and imaginary parts
(open symbols) are shown, for both parallel and perpendicular
directions. These results are compared with the theory for $\eta$ =
0.97~mPa s and $\rho$ = 1000~kg m$^{-3}$.
There is good agreement with no free parameters.}
\label{fig:water Active}
\end{figure}

We have written the response functions in Eqs.\ (\ref{alpha-par},
\ref{alpha-perp}) in a form in which the noninertial limits
($x\rightarrow 0$) are simple: $\chi_{\parallel,\perp}\rightarrow
1$. Thus, for a simple liquid in the limit $x \rightarrow 0$,
Eqs.\ (\ref{velocity-par}, \ref{velocity-perp}) reduce to the
(time-independent) Oseen tensor~\cite{brenner,Oseen}. For finite
$x$, these response functions give the dynamic Oseen tensor
\cite{Oseen,Mazur}, which are shown as the solid lines in
Fig.\ \ref{fig:water Active}, where for small $r/\delta$ the parallel and perpendicular
\emph{velocity} response (\emph{i.e.},
$-i\omega\alpha_{\parallel,\perp}$) approach $\frac{1}{4\pi\eta
r}$ and $\frac{1}{8\pi\eta r}$ for a unit force at the origin.
These then decay for $r\agt\delta$. The region of negative
response in the perpendicular case corresponds to the back-flow of
the vortex.

The response functions above represent the ensemble-average
displacements due to forces acting in the medium. These response
functions also govern the equilibrium thermal fluctuations and the
correlated fluctuations from point to point within the medium. The
relationship between thermal fluctuations and response is
described by the Fluctuation-Dissipation theorem. Specifically,
for points separated by a distance $r$ along the $\hat x$
direction,
\begin{eqnarray}
C_{\parallel,\perp}(r,\omega)=
\frac{2k_BT}{\omega}\alpha''_{\|,\perp}(r,\omega),
\label{corr}
\end{eqnarray}
where
\begin{equation}
C_\|(r,\omega)=\int_{-\infty}^\infty dt e^{i \omega t}\langle
u_x(0,0)u_x(r,t)\rangle
\end{equation}
and
\begin{equation}
C_\perp(r,\omega)=\int_{-\infty}^\infty dt e^{i \omega t}\langle
u_y(0,0)u_y(r,t)\rangle.
\end{equation}

\subsubsection{Polymer solutions}
\begin{figure}\centering
\includegraphics[width=8cm]{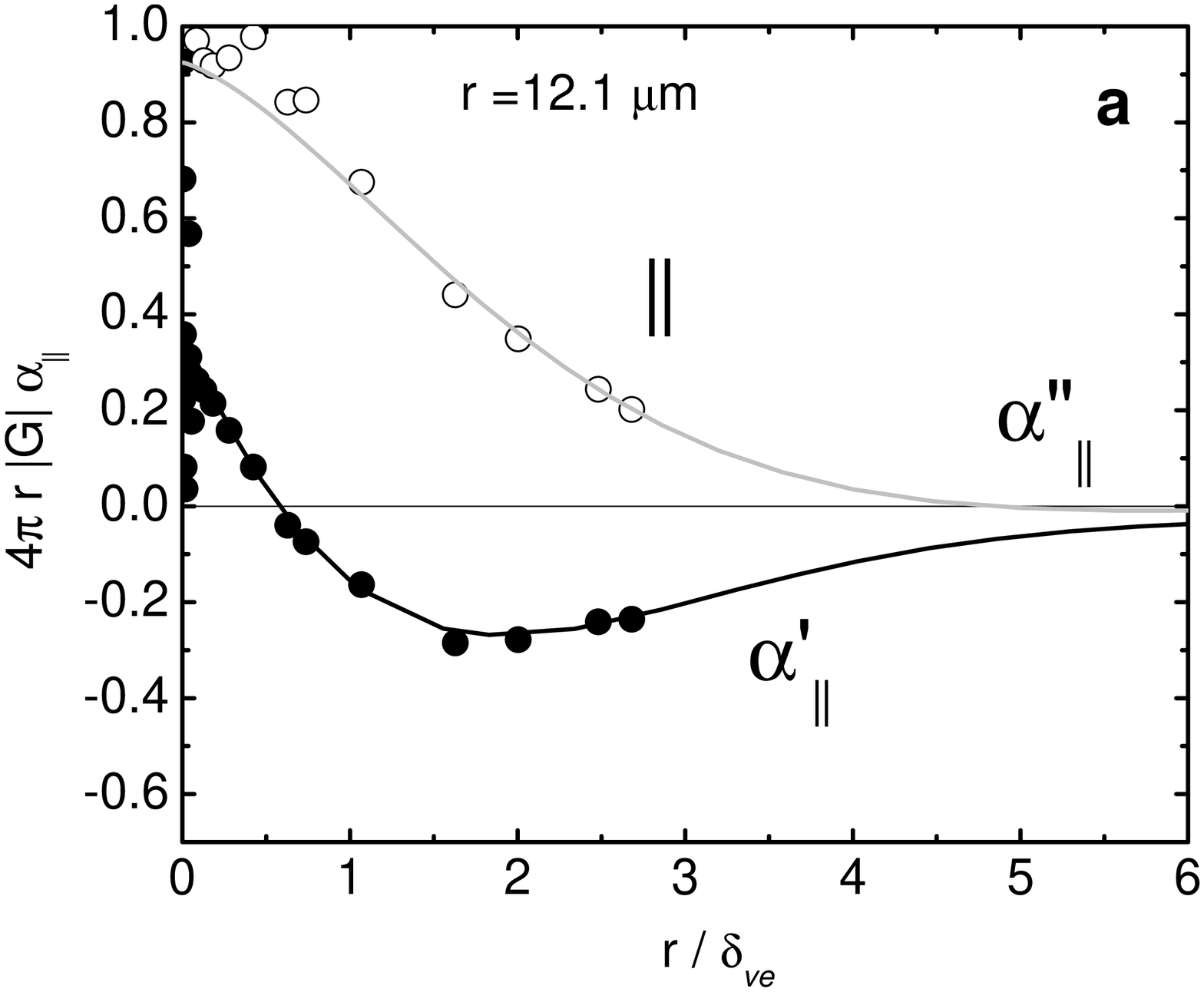}
\includegraphics[width=8cm]{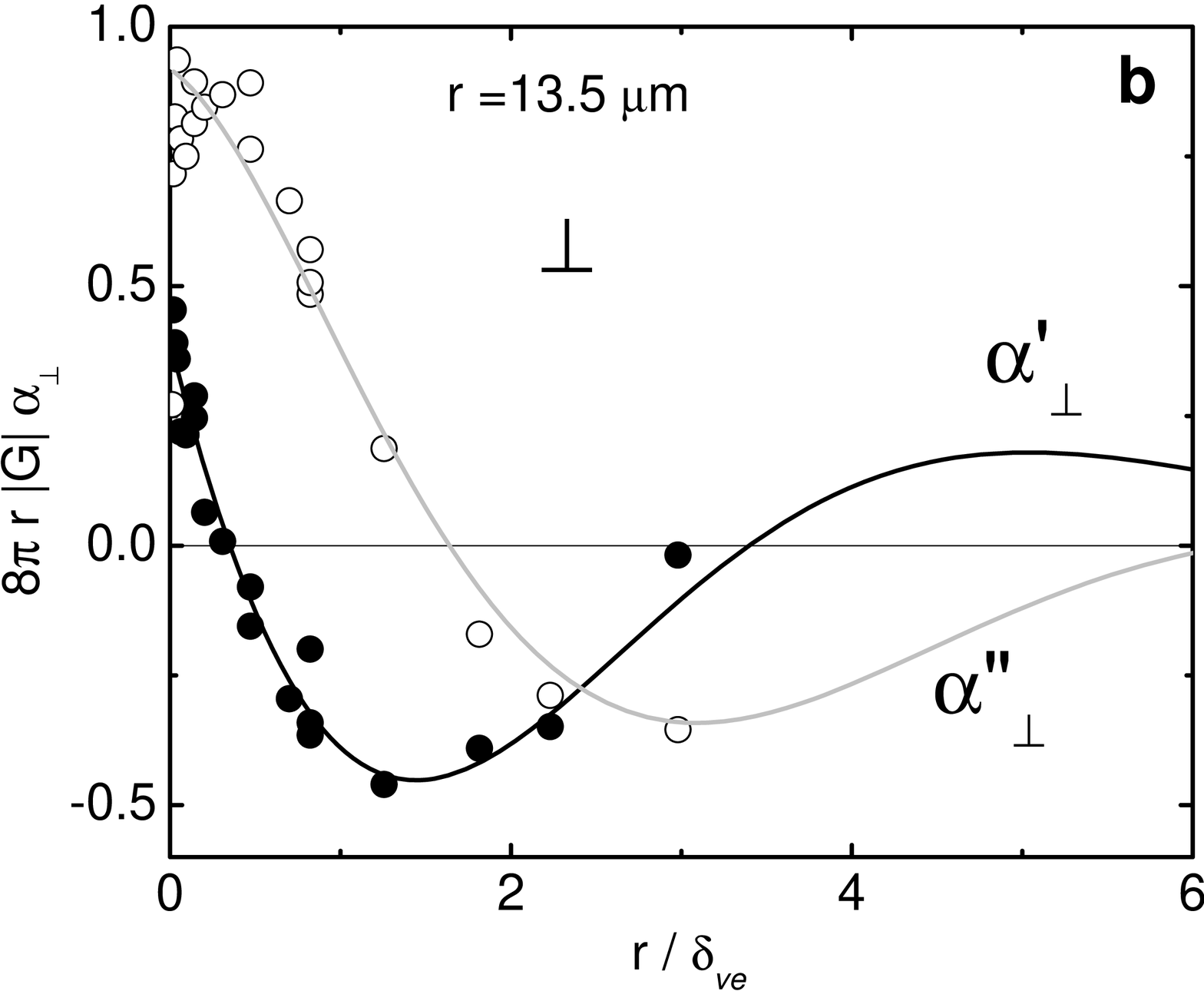}
\caption{Comparison to Eqs.\ (\ref{alpha-par}-\ref{chi_perp}) 
of both real and imaginary
parts of the normalized inter-particle response functions between
two probe particles (silica beads, $R$ = 0.58~$\mu$m), measured
with the active microrheology method for two separation distances $r$ in 1 mg/ml
entangled F-actin solutions. (a) $4 \pi r |G| \alpha_{||}(\omega)$
and (b) $8 \pi r |G| \alpha_{\bot}(\omega)$, plotted versus the
ratio of the separation distance $r$ ($r$ = 11.3~$\mu$m in
parallel and $r$ = 10.3~$\mu$m in perpendicular direction) to the
frequency-dependent viscoelastic penetration depth $\delta_{ve}$.
Here, the parameters $\bar g$ and $z$ in Eq.\ (\ref{Gomega}) were varied to
obtain simultaneous fits of all data sets to Eqs.\ (\ref{re_alpha_parr}-\ref{im_alpha_perp}),
using a single set of parameters $z$ and $\bar g$, while accounting for
the solvent (water) viscosity. Both the real (filled symbols) and
imaginary (open symbols) parts of both parallel and perpendicular
response functions are in a good agreement with Eqs.\ (\ref{re_alpha_parr}-\ref{im_alpha_perp}) with optimal parameters $\bar g$ = 0.22 $\pm$ 0.05~Pa s$^{z}$ and $z$ = 0.78
$\pm$ 0.02. The corresponding theoretical lines are shown for $z$
= 0.75 and $\bar g$ = 0.22.} \label{fig:actin Active}
\end{figure}

An experimentally pertinent illustration is given by the {\em high
frequency} complex shear modulus of a polymer solution,
\begin{equation}
G^\star(\omega)=-i \omega \eta + \bar{g}(-i\omega)^z = |G|e^{-i
\psi}\label{Gomega}
\end{equation}
which has both solvent and polymer contributions.  For the
Rouse model of flexible polymers $z=1/2$ \cite{Doi}, while
for semiflexible polymers $z=3/4$ \cite{MR97,Morse98,Gittes98}.
The latter case is shown as the solid lines in Fig.\ \ref{fig:actin Active}.
We see that the oscillatory or anti-correlated response becomes
more pronounced in viscoelastic materials.

The magnitude of the complex modulus is given by
\begin{equation}
|G|= \sqrt{ \left( \bar g \omega^z\right)^2 + \left( \omega \eta
\right)^2 + 2 \omega^{z+1} \eta \bar g \sin \left( \pi z /2
\right)},
\end{equation}
while its phase is given by
\begin{equation}
\sin \psi = \frac{\left( \omega \eta + \bar g \omega^z \sin \left(\pi z
/2 \right) \right)}{|G|}
\end{equation}
and
\begin{equation}
\quad \cos \psi = \frac{\bar g \omega^z \cos \left(\pi z /2 \right)}
{|G|} \, .
\end{equation}
It is also useful to have the following expressions for the
half-phase-angles
\begin{eqnarray}
\sin {\psi \over 2} &=& \sqrt{{1 \over 2} \left( 1 - {\bar g
\omega^z \cos \left(\pi z /2  \right) \over |G|} \right)} \\
\cos {\psi \over 2} &=& \sqrt{{1 \over 2}\left(1+ {\bar g \omega^z
\cos \left(\pi z /2  \right)  \over |G|} \right) } \quad .
\end{eqnarray}
\begin{widetext}

We define the real parameter $ \displaystyle \beta = r  \sqrt{\rho
\omega^2 /|G| }$ and use the definitions, Eqns.
(\ref{chi_parr},\ref{chi_perp}) to obtain the following compact
expressions for the response functions ($\alpha_\perp, \alpha_\|$)
which can be expanded using the compound angle formulae and the
definitions above. The real and imaginary parts of the parallel
response are given by
\begin{equation}
4\pi|G|r\alpha'_\| (r,\omega) = \frac{2}{\beta^2} \left \{ e^{-
\beta \sin {\psi \over 2}} \left[ \cos \left(\beta \cos {\psi
\over 2} \right) + \beta \sin \left( {\psi \over 2} + \beta \cos
{\psi \over 2} \right)  \right] - 1 \right\} \label{re_alpha_parr}
\end{equation}
and
\begin{equation} 4\pi|G|r\alpha''_\| (r,\omega)=
\frac{2}{\beta^2} e^{- \beta \sin {\psi \over 2}} \left \{ \sin
\left(\beta  \cos {\psi \over 2} \right) - \beta \cos \left( {\psi
\over 2} + \beta \cos {\psi \over 2} \right) \right\},
\label{im_alpha_parr}
\end{equation}
while the corresponding expressions for the perpendicular response
are given by
\begin{equation}
8\pi|G|r\alpha'_\perp (r,\omega)= \frac{2}{\beta^2} \left \{ 1 -
e^{- \beta \sin {\psi \over 2}} \left[\cos \left(\beta \cos {\psi
\over 2} \right) + \beta \sin \left( {\psi \over 2} + \beta \cos
{\psi \over 2} \right) - \beta^2 \cos \left(\psi+ \beta \cos {\psi
\over 2}\right) \right] \right\} \label{re_alpha_perp}
\end{equation}
and
\begin{equation} 8\pi|G|r\alpha''_\perp (r,\omega)=
\frac{2}{\beta^2} e^{- \beta \sin {\psi \over 2}} \left \{ - \sin
\left(\beta \cos {\psi \over 2} \right) + \beta \cos \left( {\psi
\over 2} + \beta \cos {\psi \over 2} \right) + \beta^2 \sin
\left(\psi+ \beta \cos {\psi \over 2}\right) \right\}.
\label{im_alpha_perp}
\end{equation}

The imaginary part of the response functions will be used to
calculate the correlation functions, Eq.\ (\ref{corr}), used for
analysis of the passive experiments whilst the real part of the
response functions will be used for comparison with the active
experiments.

We can simplify the expressions above in the limit that the
polymer contribution to the viscoelasticity dominates the shear
modulus. We then obtain the simple scaling form $G^\star(\omega)
\simeq \bar{g}(-i\omega)^z$\cite{NoteWater} giving
$|G|=\bar{g}\left|\omega \right|^z, \psi= \pi z/2$. Further
simplification of Eq.\ (\ref{corr}) using the expressions in Eqs.\
(\ref{im_alpha_perp},\ref{im_alpha_parr}) and definitions in Eqs.\
(\ref{chi_parr},\ref{chi_perp}) leads, \emph{e.g.}, to
\begin{equation}
C_\parallel(r,\omega)= \frac{k_BT}{2\pi \omega|G|
r}\Big\{\frac{2}{\beta^2}e^{-\sin\left(\tfrac{z\pi}{4}\right)\beta}
\Big[ \Big(1+\sin\left(\tfrac{z\pi}{4}\right)\beta\Big)
\sin\left[\cos\left(\tfrac{z\pi}{4}\right)\beta\right]
-\cos\left(\tfrac{z\pi}{4}\right)\beta
\cos\left[\cos\left(\tfrac{z\pi}{4}\right)\beta\right]
\Big]\Big\},
\end{equation}
where
$
\beta=r\sqrt{\rho\omega^2/|G|}
$ 
characterizes the overall decay of stress due to inertia. This decay corresponds
to super-diffusive propagation of stress for viscoelastic media with 
$G\sim\omega^{z}$ and $z<1$, since the response is limited to a spatial 
range that grows with time as $t^{(2-z)/2}$.
\end{widetext}

\begin{figure}
\centering
\includegraphics[width=8cm]{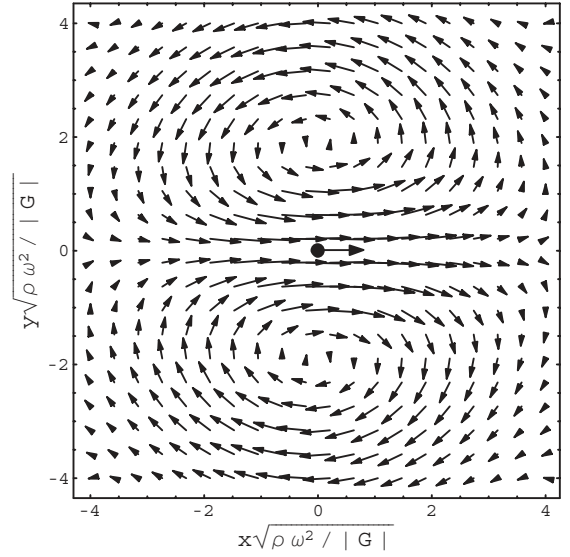}
\caption{The displacement field displays a clear vortex-like
structure. Here, a force in the $\hat x$ direction is applied at
the origin (as shown by the filled circle and arrow). Distances
are shown in units of the penetration depth
$\delta=\sqrt{|G|/(\rho\omega^2)}$. This example has been
calculated for the Rouse model with $z=1/2$.} \label{fig:vortex}
\end{figure}

The resulting displacement field, exhibiting the vortex pattern,
is shown in Fig.\ \ref{fig:vortex} for a point force at the origin
pointed along the $x$-axis. This flow pattern exhibits specific
inversion symmetries: $v_x$ ($v_y$) is symmetric (antisymmetric)
for either $x\rightarrow -x$ or $y\rightarrow -y$, as can be seen
by the fact that the (linear) response must everywhere reverse if
the direction of the force is reversed.

\subsection{Velocity autocorrelations and the \textit{long time tail}}

The self-sustaining back-flow represented in Fig.\
\ref{fig:vortex} gives rise to long-lived correlations that, for
instance, affect the crossover from ballistic to diffusive motion
of a particle in a liquid. For a simple liquid, the fluid velocity
(auto)correlations $\langle \vec v(0,t)\cdot\vec v(0,0)\rangle$
decay proportional to $\sim|t|^{-3/2}$. This is known as the
\emph{long time tail} \cite{theor_tails,Alder,exper_tails}. For a
viscoelastic fluid, stress propagation is faster than diffusive,
resulting in a more rapid decay of velocity correlations. The
decay is, however, still algebraic.

\subsubsection{Simple liquids} For a simple liquid, Eq.\
(\ref{eq:linNS})
means that
\begin{equation}
\left(-i\omega\rho+\eta k^2\right)v_i=\left(\delta_{ij}-\hat
k_i\hat k_j\right)f_j
\end{equation}
for the Fourier transforms. This gives the response \emph{in
velocity} $v_i$ to a force component $f_j$ that is
(thermodynamically) conjugate to a displacement $u_j$. We denote
this response function by $\chi_{v_i u_j}$, where
\begin{equation}
\chi_{v_i u_j}(\vec k,\omega)=\frac{\left(\delta_{ij}-\hat k_i\hat
k_j\right)}{\eta k^2-i\omega\rho}.
\end{equation}

The fluctuation-dissipation theorem then tells us that
\begin{equation}
\chi_{v_i u_j}(\vec r,t) =-\frac{1}{kT}\frac{d}{dt}\langle
v_i(\vec r,t)u_j(0,0)\rangle,
\end{equation}
where this is valid only for $t>0$ because of causality in the
response. The correlation function is, however, defined for both
positive and negative times.

Due to translation invariance in time, the correlation function
$\langle v_i(\vec r,t+t')u_j(0,t')\rangle$ must be independent of
$t'$. Thus,
\begin{eqnarray}
0&=&\frac{d}{dt'}\langle v_i(\vec r,t+t')u_j(0,t')\rangle\\
&=&\langle \dot v_i(\vec r,t+t')u_j(0,t')\rangle
+\langle v_i(\vec r,t+t')v_j(0,t')\rangle,\nonumber
\end{eqnarray}
which also means that
\begin{equation}
kT\chi_{v_i u_j}(\vec r,t)
\rangle\nonumber\\
=\langle v_i(\vec r,t)v_j(0,0)\rangle.
\end{equation}
Again, this is valid only for $t>0$ because of causality in the
response. Ultimately, however, we are interested in the
autocorrelation function $\langle \vec v\cdot\vec v\rangle=\langle
v_i v_i\rangle$, for $r\rightarrow0$ above. In this case, the
correlation function is manifestly symmetric in $t$. Thus,
\begin{equation}
\langle v_i(\vec r\rightarrow0,t)v_i(0,0)\rangle =kT\chi_{v_i
u_i}(\vec r\rightarrow0,|t|).
\end{equation}

We first calculate $\chi_{v_i u_j}\left(\vec k,t\right)$, since
the limit $\chi_{v_i u_j}\left(\vec r\rightarrow0,t\right)$ can be
obtained from this simply by integrating over all $\vec k$.
\begin{eqnarray}
\chi_{v_i u_j}(\vec k,t)&=&\left(\delta_{ij}-\hat k_i\hat
k_j\right)\int\frac{d\omega}{2\pi}
e^{-i\omega t}\frac{1}{\eta k^2-i\omega\rho}\nonumber\\
&=& \frac{\left(\delta_{ij}-\hat k_i\hat k_j\right)}{-2\pi
i\rho}\int e^{-i\omega t}\frac{d\omega}{\omega+i\nu k^2},\nonumber
\end{eqnarray}
where $\nu=\eta/\rho$. But, the last integral can only depend on
the combination $t\nu k^2$, since we can replace $\omega$ by
$\zeta\nu k^2$, where $\zeta$ is dimensionless. Specifically,
\begin{eqnarray}
\chi_{v_i u_j}(\vec k,t)&=&\frac{\left(\delta_{ij}-\hat k_i\hat
k_j\right)}{-2\pi i\rho}\int
e^{-i\zeta \nu k^2 t}\frac{d\zeta}{\zeta+i}\nonumber\\
&=&\frac{\left(\delta_{ij}-\hat k_i\hat k_j\right)}{\rho}e^{-\nu
k^2 t}
\end{eqnarray}
for $t>0$. Otherwise, the result is zero. This integral can be
done by integration along a closed contour containing the real
line in either the upper half-plane for $t<0$, or the lower half
plane for $t>0$.

Finally, to get the limit
\begin{equation}
\chi_{v_i u_i}(\vec r\rightarrow0,t)
\end{equation}
we simply integrate:
\begin{eqnarray}
\chi_{v_i u_i}(\vec r\rightarrow0,t) &=&\int\frac{d^3 k}{(2\pi)^3}
\frac{\left(\delta_{ii}-\hat k_i\hat k_i\right)}{\rho}e^{-\nu k^2 t}\nonumber\\
&=&\frac{2}{\rho}\left(4 \pi\nu t\right)^{-3/2}.
\end{eqnarray}
Again, this is only for $t>0$. Thus,
\begin{equation}
\langle \vec v(\vec r\rightarrow0,t)\cdot\vec v(0,0)\rangle
=\frac{2kT}{\rho}\left(4 \pi\nu |t|\right)^{-3/2}.
\end{equation}

\subsubsection{Viscoelastic media} For viscoelastic media, the
calculation is similar, except that
\begin{eqnarray}
\chi_{v_i u_j}(\vec k,\omega)&=&-i\omega
\chi_{u_i u_j}(\vec k,\omega)\nonumber\\
&=&\chi_{vu}(\vec k,\omega)\left(\delta_{ij}-\hat k_i\hat
k_j\right),
\end{eqnarray}
where $\chi_{u_i u_j}=\alpha_{ij}$, and where we have defined
\begin{equation}
\chi_{vu}(\vec
k,\omega)=\frac{1}{\rho}\left(\frac{i\omega}{\omega^2-G^*k^2/\rho}\right)
\label{chi_ve_freq}\end{equation} for simplicity. As above, all
the singularities in this must lie in the lower half-plane in
order for the response function to be causal.

Evaluation of the inverse Fourier transform of Eq.\
(\ref{chi_ve_freq}) to obtain $\chi_{vu}(\vec k,t)$ can be done by
use of the Mittag-Leffler functions. The Mittag-Leffler
functions~\cite{podlubny,erdelyi}, $E_\alpha(z)$, which are entire
functions parameterized by a continuous parameter $\alpha$ can be
defined by the power series
\begin{equation}
E_\alpha  (z) = \sum_{k=0}^\infty {z^k \over \Gamma(1 + \alpha k)}
\quad ; \quad \alpha > 0 \quad . \label{mittag}
\end{equation}
Note $E_1(x) = \exp (x)$.  Straightforward
manipulation~\cite{podlubny} of the definition, Eq.
(\ref{mittag}) show that their causal Fourier transforms are given
by
\begin{equation}
\int_{-\infty}^\infty dt e^{i \omega t} \Theta(t) E_\alpha(-a
t^\alpha) = { i \omega \over \omega^2 - a \left(-i
\omega\right)^{2-\alpha}}
\end{equation} where
$\Theta(t)$ is the Heaviside step function. Performing the inverse
Fourier transform, we obtain
\begin{equation}
\chi_{vu}(\vec k,t) = E_{2-z}\left[- (\bar{g} k^2 / \rho) t^{2-z}
\right] \quad ; \quad t > 0
\end{equation}
for $G^*=\bar{g} (- i \omega)^z$.
The asymptotic expansions for the Mittag-Leffler
functions~\cite{podlubny,paris} are
\begin{equation}
E_\alpha (- z) = \left\{ \begin{array}{c} \displaystyle 1 - {z
\over \Gamma(1+\alpha)} + O(z^2)\,, \quad z \ll 1 \\
\displaystyle {z^{-1} \over \Gamma(1-\alpha)} + O(z^{-2})\,, \quad
z \gg 1 \end{array} \right.
\end{equation}

Thus, the velocity correlation function  is given by
\begin{eqnarray} 
\langle \vec v(\vec r\rightarrow0,t)\cdot\vec
v(0,0)\rangle = && \\ \frac{2kT}{\rho} { 4 \pi \over (2
\pi)^3} \int_0^{1/a}&&\!\!\!\!\!\!\!\!\! k^2 dk E_{2-z}\left[- (\bar{g} k^2 / \rho)
|t|^{2-z} \right]\nonumber 
\end{eqnarray}
From the asymptotic properties of $E_\alpha(t)$, it is clear that
the integral does not converge at $k \rightarrow \infty$
necessitating a finite cut-off (which we choose as the size of the
probe particle). We can express the correlation function as
\begin{eqnarray} 
\langle \vec v(0,t)\cdot\vec
v(0,0)\rangle &=& |t|^{-\frac{3}{2}{(2-z)}} \frac{2kT}{\rho} \left( {\rho
\over \bar{g}} \right)^{3/2}{ 2-z \over (2 \pi)^2} I_{2-z}
\nonumber \\ I_\alpha &=& \int_0^{\Lambda_\alpha} dx
x^{3\alpha/2-1}  E_{\alpha}\left(- x^{\alpha} \right)
\end{eqnarray}
and $\Lambda_\alpha= (|t|\bar{g}/\rho a^2)^{1/\alpha}$.

We can get an approximation to  the value of $I_\alpha$ by
splitting the integral into two sections
$$I_\alpha = \int_0^1 dx E_\alpha^{\ll}(-x^\alpha) + \int_1^{\Lambda_\alpha} dx E_\alpha^{\gg}(-x^\alpha)$$ using the two asymptotic forms for $E_\alpha(t)$.

We then obtain finally the expression
\begin{equation}
\langle \vec v(\vec r\rightarrow0,t)\cdot\vec v(0,0)\rangle \simeq
C_1 |t|^{-3(2-z)/2} + C_2 |t|^{-(5-3z)/2}\label{viscoelLLT}
\end{equation}
where
\begin{eqnarray}
C_1&&=\frac{4kT}{(2 \pi)^2 \rho} \left( {\rho \over \bar{g}}
\right)^{3/2}\times\\
&&\frac{5 \Gamma(3-z)\Gamma(z-1) - 3 \Gamma(z-1)-15
\Gamma(3-z)}{15 \Gamma(3-z)\Gamma(z-1)}\nonumber
\end{eqnarray}
and
\begin{equation}
C_2= \frac{4kT}{(2 \pi)^2 a \rho} \left( {\rho \over \bar{g}}
\right)^{}{1\over  \Gamma(z-1)}.
\end{equation}
Here, the dominant first term in Eq.\ (\ref{viscoelLLT}) corresponds to a
faster asymptotic decay of correlations for $z<1$ than for simple liquids. 
This is a direct consequence of the super-diffusive propagation of stress in this
case. Also, it is interesting to note that the second term in Eq.\ (\ref{viscoelLLT})
strictly vanishes in the $z\rightarrow 1$ limit of simple liquids.

\section{MATERIALS AND METHODS}

Simple liquids: For our experiments we used two Newtonian fluids
with different viscosities $\eta$ and mass densities $\rho$,
namely water ($\eta$ = 0.969~mPas and $\rho$ = 1000~kg$m^{-3}$)
and a $(1:1~ v/v)$ water/glycerol mixture ($\eta$ = 6.9~mPas and
$\rho$ = 1150~kg$m^{-3}$).

Viscoelastic fluids: We performed experiments with two different
viscoelastic fluids, worm-like micelle solutions and solutions of
the cytoskeletal biopolymer F-actin. Worm-like micelles were
prepared by self-assembly of cetylpyridinium chloride (CPyCl) in
brine (0.5M~NaCl) with sodium salicylate (NaSal) as counterions,
with a molar ratio of Sal/CPy = 0.5. Three different
concentrations of worm-like micelles were used: $c_{m}$ = 0.5\%,1\%
and 2\%. Worm-like micelles behave essentially like linear
flexible polymers with an average diameter of about 3~nm, a
persistence length of about 10~nm, and contour lengths of several
$\mu$m \cite{Berret}. F-actin was polymerized from monomeric actin
(G-actin) isolated from rabbit skeletal muscle according to a
standard recipe ~\cite{Pardee}. G-actin was mixed with silica
beads in polymerization buffer [2~mM HEPES, 2~mM MgCl$_{2}$, 50~mM
KCl, 1~mM Na$_{2}$ATPa, and 1~mM EGTA, pH 7] and incubated for 1
hour. Entangled actin solutions were used as a model system for
semiflexible polymer solutions. Experiments were done at
concentrations of c = 0.5 and 1~mg/ml.

\subsection{Experimental methods}\label{sec:Meth}

Details of the experimental set-up can be found in \cite{Allersma,
setup maryam, Mizuno active, MizunoTechnical}. Briefly, we have used a custom-built
inverted microscope that includes a pair of optical traps formed
by two focused laser beams of different wavelengths ($\lambda$ =
1064 nm, ND:YVO$_{4}$, Compass, Coherent, Santa Clara, CA, USA)
and $\lambda$ = 830~nm (diode laser, CW, IQ1C140, Laser 2000). The
optical traps fulfil two functions: (i) They confine the particles
around two well-defined positions and at the same time detect
particle displacements with high temporal and spatial resolution
in the passive method. (ii) In the active method one trap is used
to apply a sinusoidally varying force to one particle, while the
response of the other particle is detected in the second trap. In
the passive method, a pair of silica beads (Van't Hoff Laboratory,
Utrecht University, Utrecht, Netherlands) of various radii (R =
0.25~ $\mu$m $\pm$5\%, 0.580 $\mu$m $\pm$5\%, 1.28$\mu$m $\pm$5\%
and 2.5 $\mu$m$\pm$5\%) were weakly trapped (trap stiffness
between 2~$\mu$N/m and 5~$\mu$N/m, where larger particles required
the higher laser intensities to avoid shot noise. Transmitted
laser light was imaged onto two quadrant photo diodes, such that
particle displacements and in the $x$ and $y$ directions were
detected interferometrically \cite{detection Gittes}. A
specialized silicon PIN photodiode (YAG444-4A, Perkin Elmer,
Vaudreuil, Canada), operated at a reverse bias of 110~V, was used
in order to extend the frequency range up to 100~kHz for the
1064~nm laser \cite{Erwin detection}. The 830~nm laser was
detected by a standard silicon PIN photodiode operated at a
reverse bias of 15~V (Spot9-DMI, UDT, Hawthorne, CA). Amplified
outputs were digitized at 195~kHz (A/D interface specs ) and
further processed in Labview (National Instruments, Austin, TX,
USA). Output voltages were converted to actual displacements using
Lorentzian fits to power spectral densities (PSD) as described in
\cite{signal and noise}. In the case of water, calibration was
done on the beads that were used in the experiments, while for the
viscoelastic solutions and the more viscous liquid, calibrations
were done in water with beads from the same batch.

In the active
method, the 1064~nm laser was used to oscillate one particle,
while the 830~nm laser was used for detection of the second
particle at a separation distance $r$. The driving laser was
deflected through an Acousto-Optical Deflector (AOD) (TeO$_{2}$,
Model DTD 276HB6, IntraAction, Bellwood, Illinois), using a
voltage-controlled oscillator (VCO) (DRF.40, AA OPTO-ELECTRONIC,
Orsay, France). The force applied to the driven particle was
calibrated by measuring the PSD of the Brownian motion of a
particle of the same size trapped in water with the same laser
power \cite{signal and noise}. The output signal from the QPD
detecting the second laser was fed into a lock-in amplifier
(SR830, Stanford Research Systems, Sunnyvale, CA, USA) to obtain
amplitude and phase of particle response. All experiments were
done in sample chambers made from a glass slide and a cover slip
with about 140~$\mu$m inner height, with the particles at at least
25~$\mu$m distance from both surfaces. The lab temperature was
stabilized at T = $21.5~^{\circ}$C.

\section{DATA ANALYSIS}

\begin{figure}
\centering
\includegraphics[width=240pt]{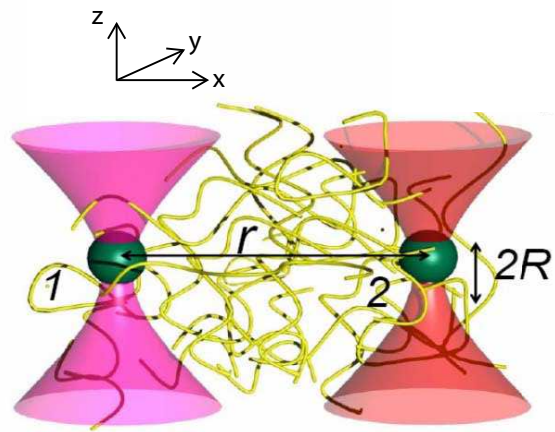}
\caption{(Color online) Schematic sketch of the experiment. A pair
of silica beads (radius $R$) is trapped by a pair of laser traps
at a separation distance $r$. In the passive method, the position
fluctuations of each particle in the $x$ and $y$ directions are
simultaneously detected with quadrant photodiodes and the
displacement cross-correlations are measured parallel and
perpendicular to the line connecting the centers of the two beads.
In the active method, one of the beads is oscillated by rapidly
moving one laser trap either in $x$ or in $y$ direction and the
resulting motion of the other particle is measured in $x$ and in
$y$ direction. Laser intensity was adjusted to result in the trap
stiffness of typically between (2$\mu N/m$ and 5$\mu N/m$) for
passive measurements.} \label{fig:sketch}
\end{figure}

In both the active and passive methods we calculate the linear
complex response function $\alpha$  defined by $u(\omega) =
\alpha(\omega)\times F(\omega)$, where $F(\omega)$ is the applied force.
Linear response applies by definition in the passive method, and
in the active method the particle displacements $u(\omega)$ were
kept sufficiently small. Again, we consider separately real,
$\alpha'(\omega)$, and imaginary parts, $\alpha''(\omega)$, of the
response. In all our experiments, as sketched in Fig.\ \ref{fig:sketch},
the coordinate system was chosen
in such a way that $x$ is parallel to the line connecting the
centers of the two particles ($||$) and $y$ perpendicular
($\bot$)to that. The inter-particle response functions along these
two directions were used to determine the flow field.

The displacement $u^{(1)}_x(\omega)$ of particle 1 in the $x$
direction is related to the force $F^{(2)}_x$ acting on particle 2
according to $u^{(1)}_x (\omega)=\alpha_{||}(\omega)\times
F^{(2)}_x(\omega)$. Similarly, the perpendicular response function
was derived from $u^{(1)}_y (\omega)=\alpha_{\bot}(\omega)\times
F^{(2)}_y(\omega)$. The single-particle response functions for
each $x$ and $y$ directions are defined as $u^{(1)}_{x,y}(\omega)
= \alpha_{\rm auto}(\omega)\times F^{(1)}_x(\omega)$. For
homogeneous, isotropic media, the two functions
$\alpha_{||,\bot}(\omega)$ completely characterize the linear
response at any point in the medium due to a force at another
point. The displacement response functions
$\alpha_{||,\bot}(\omega)$ determine both position and velocity
 response $-i\omega\alpha_{||,\bot}(\omega)$.

In the passive approach, the medium fluctuates in equilibrium,
and the only forces on the particles are thermal/Brownian forces.
Therefore the fluctuation-dissipation theorem (FDT) of statistical
mechanics \cite{Landau stat} relates the response of the medium to
the displacement correlation functions. For two particles, these
correlation functions are the cross-correlated displacement
fluctuations: $\langle u^{(1)}_{x}(t)u^{(2)}_{x}(0)\rangle$ and
$\langle u^{(1)}_{y}(t)u^{(2)}_{y}(0)\rangle$. We used Fast
Fourier Transforms (FFT) to calculate displacement
cross-correlation functions in frequency space and obtained the
imaginary parts of the complex inter-particle response functions
$\alpha''_{||,\bot}(\omega)$ via the FDT:
\begin{equation}
\alpha''_{||}(\omega)=\frac{\omega\int\langle
u^{(1)}_{x}(t)u^{(2)}_{x}(0)\rangle e^{i\omega t}dt}{2kT}
\end{equation}
and
\begin{equation}
\alpha''_\bot(\omega)=\frac{\omega\int\langle
u^{(1)}_{y}(t)u^{(2)}_{y}(0)\rangle e^{i\omega t} dt}{2kT},
\end{equation}
where k is the Boltzman constant and T is the controlled
laboratory temperature. The real parts of inter-particle response
functions $\alpha'_{||,\bot}(\omega)$  were obtained by a
Kramers-Kronig integral:
\begin{eqnarray}
\alpha'_{||,\bot}(\omega)&=&\frac{2}{\pi}P\int^{\infty}_0
\frac{\zeta\alpha''_{||,\bot}(\omega)}{\zeta^{2}-\omega^{2}}\\
&=&
\frac{2}{\pi}\int^{\infty}_0\cos(t\omega)\int^{\infty}_0
\alpha''_{||,\bot}(\zeta)\sin(t\zeta)d\zeta,\nonumber
\end{eqnarray}
where P denotes a principal-value integral \cite{MR97}. The high
frequency cut-off of the Kramers-Kroning integral limits the
frequency range of the calculated $\alpha'_{||,\bot}(\omega)$
\cite{setup maryam}. We also used the active method to obtain both
real and imaginary parts of the response
functions with 100~kHz
bandwidth, as in Refs.\ \cite{Mizuno active,MizunoTechnical}.
Here the lock-in amplifier
provides directly in-phase (real part) and out-of-phase (imaginary
part) response of the second particle. The measurements were done
over a grid of driving frequencies.

\section{RESULTS}
\subsection{Simple liquids }

In the low-frequency limit, where fluid inertia can be neglected,
the inter-particle response functions are inversely related to the
shear modulus of the medium \cite{Mason,MR97,2fluid}. For a
simple viscous fluid, the response functions in this limit are
given by:
\begin{equation} \alpha_{||}= 2\alpha_{\perp}=
\frac{i}{4 \pi r \omega \eta}
\end{equation}
where $r$ is the separation distance between the two particles and
$\eta$ is the viscosity. These relations also shows (via the FDT)
a $1/r$ dependence of the Fourier transform of the displacement
cross-correlation functions:
\begin{equation}
S_{||}=\int\langle u^{(1)}_{x}(t)u^{(2)}_{x}(0)\rangle e^{i \omega
t}dt
\end{equation}
and
\begin{equation}
S_{\perp}=\int\langle
u^{(1)}_{y}(t)u^{(2)}_{y}(0)\rangle e^{i \omega t}dt
\end{equation}

\begin{figure}
\includegraphics[width=8cm]{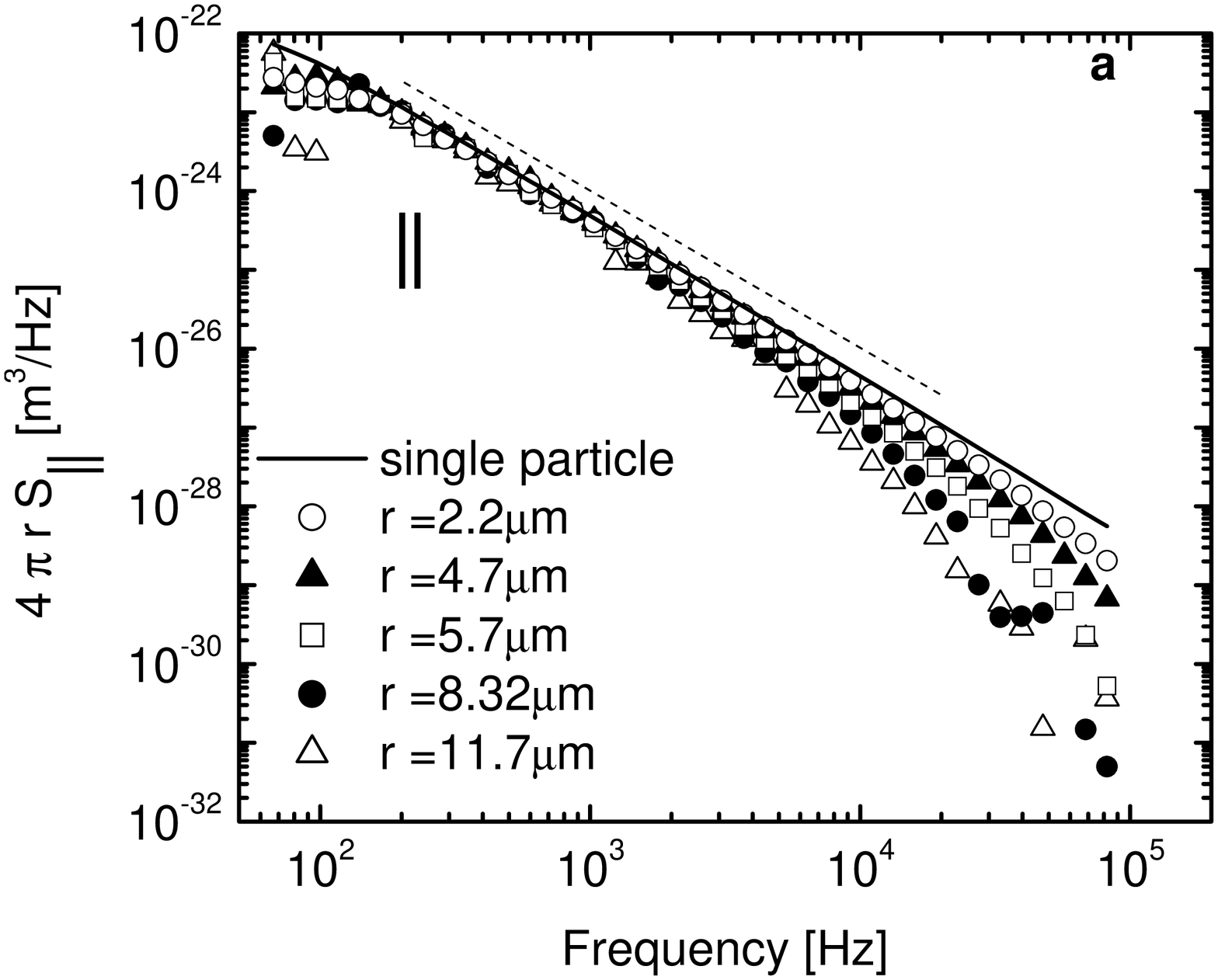}
\includegraphics[width=8cm]{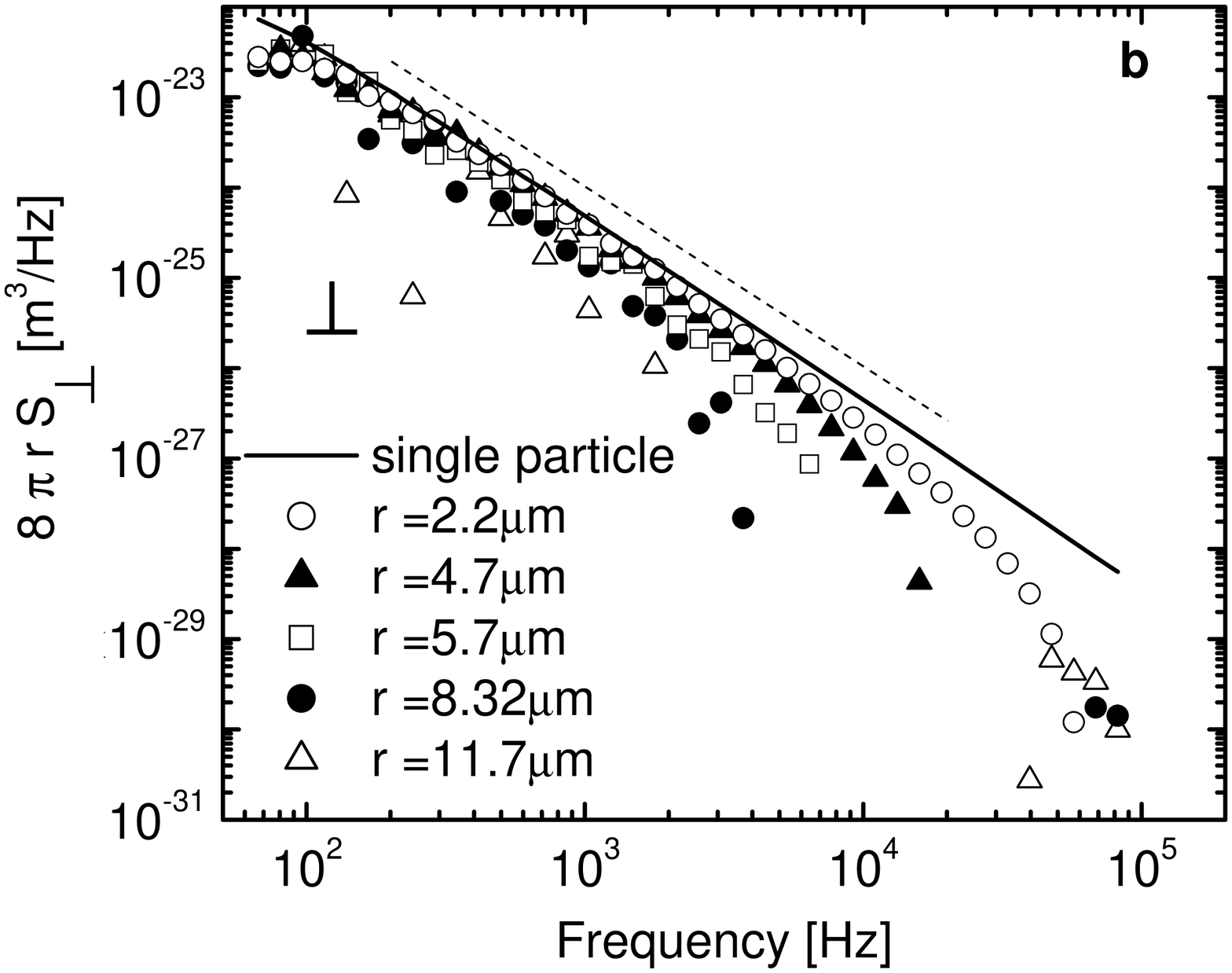}
\caption{Normalized displacement cross-correlation functions (a)
$4 \pi r S_{||}$ and (b) $8 \pi r S_{\bot}$ of two probe particles
(silica beads, $R$ = 0.58 $\mu$m) in water versus frequency ($f =
\omega/2\pi$), compared for different separation distances $r$
(symbols). In both (a) and (b), the solid line is the auto-correlation 
function of a single particle, normalized by $6 \pi R$. The single-particle
motion agrees well with the expected frequency dependence of a 
single Brownian particle (slope of -2), as indicated by the dashed lines.} 
\label{fig:water spectra}
\end{figure}

In Fig.\ \ref{fig:water spectra}, displacement cross-correlation functions for
particle pairs in water, normalized to compensate for the distance
dependence ($4 \pi r S_{||}$ and $8 \pi r S_{\bot}$ ), are plotted
versus frequency. For comparison a single-particle displacement
auto-correlation function, normalized for the bead size dependence
($6 \pi R S_{\rm auto}$), is plotted as a solid line, for a
particle radius of $R = 0.58~\mu m$. The auto-correlation function
agrees well with the power-law slope of -2 
up to nearly 100~kHz. This slope is expected from the high-frequency
limit of the Lorentzian shape of the power spectral density (=
Fourier Transform of the displacement autocorrelation function) of
the displacements of a harmonically confined Brownian particle in
a viscous fluid \cite{signal and noise}. The effect of fluid
inertia is evident as a deviation from this power law in the
displacement cross-correlation functions. A systematically $r$
dependent decrease of the cross-correlations is apparent at high
frequencies for separations r ranging from $2.2~\mu m$ to
$11.7~\mu m$. The faster decrease of cross-correlations is a
manifestation of the finite velocity at which stress propagates
into the medium. For larger $r$, the data show that the decrease
begins at a lower frequency, because it takes longer for stress to
propagate further. A comparison of Figs.\ \ref{fig:water spectra}a and b shows that the
decrease of the cross-correlation is, at the same separation
distance, more pronounced in the perpendicular channel than in the
parallel channel. This is due to the fact that in the vortex-like
flow pattern of Fig.\ \ref{fig:vortex} there is a region of fluid motion in the opposite direction to the applied force. For $r > 5~\mu m$ (open squares), the
cross-correlations become negative in the observed frequency
window (not shown in the log-log plot). At still higher
frequencies, the displacement cross-correlation functions again
become positive (Fig.\ \ref{fig:water spectra}b), which is visible for the larger
separations, $r = 8.3~\mu m$ and $r = 11.7~\mu m$, consistent with
the expected oscillation in the displacement cross-correlation
functions in the frequency domain. This effect becomes more
pronounced in viscoelastic media

\begin{figure}
\centering
\includegraphics[width=8cm]{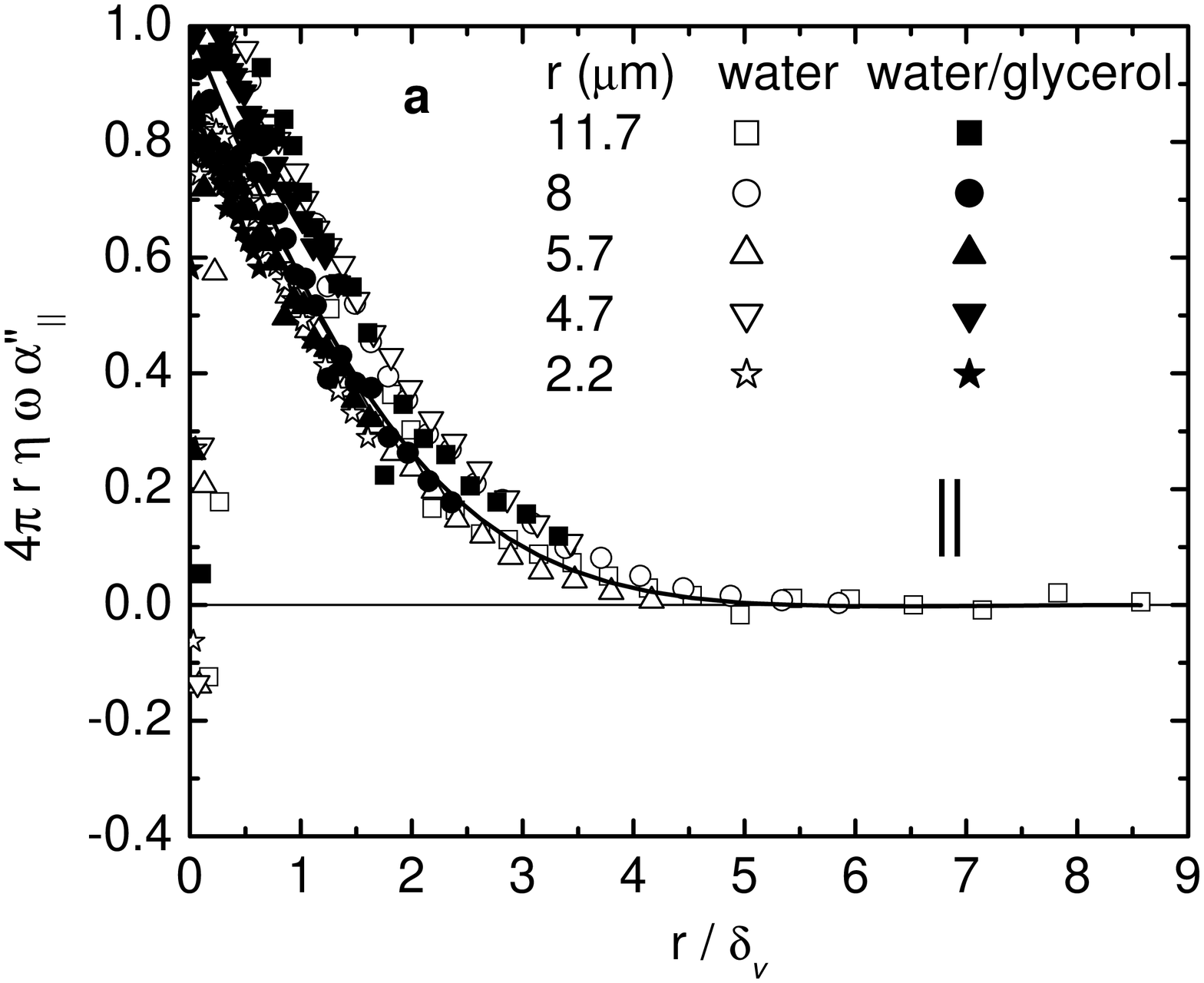}
\includegraphics[width=8cm]{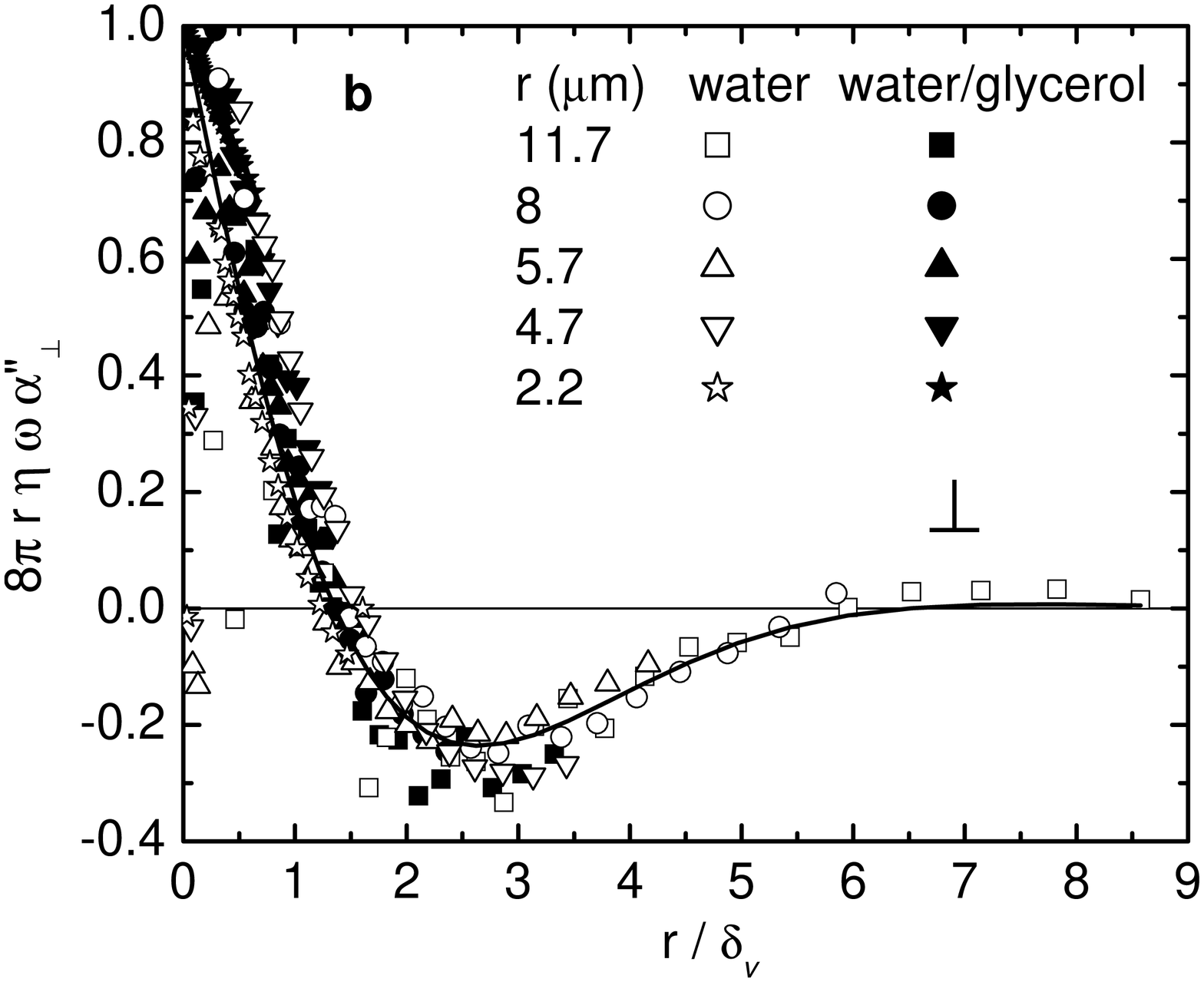}
\caption{Normalized imaginary parts of inter-particle response
functions between two probe particles (silica beads, $R$ = 0.58
$\mu$m) measured with the passive method, (a) $4 \pi r \eta \omega
\alpha''_{||}$ in the parallel direction and (b) $8 \pi r \eta
\omega \alpha''_{\bot}$ in the perpendicular direction, plotted
versus the ratio of the separation distance $r$ (fixed for a given
bead pair) to the frequency-dependent viscous penetration depth
$\delta_{v}$, in water (open symbols, $\eta$ = 0.969~mPa s) and in
water/glycerol (filled symbols, $\eta$ = 6.9~mPa s). Solid lines
are Oseen's predictions for a simple liquid with no adjustable
parameters. Data is only plotted for $\omega >$ 200 rad/s and for
$\omega <$ 2krad/s one in every 5 data points is shown.}
\label{fig:water normalized}
\end{figure}

The spatial and temporal propagation of the inertial vortex in a
general viscoelastic medium is characterized by Eqs.\ (\ref{alpha-par}-\ref{chi_perp})
\cite{Oseen, maryam, liverpool}. For simple liquids, A = $\eta$
(viscosity) and z = 1. In Figs.\ \ref{fig:water normalized}a and b, we compare the imaginary
parts of the normalized inter-particle response functions $4 \pi
\eta r \omega \alpha''_{||}(\omega)$ (Fig.\ \ref{fig:water normalized}a) and $8 \pi \eta r
\omega \alpha''_{\bot}(\omega)$ (Fig.\ \ref{fig:water normalized}b) for water and a (1:1
v/v) water/glycerol mixture. In order to collapse all data onto a
single master curve, as suggested by Eqs.\ (\ref{alpha-par}-\ref{chi_perp}),
we have plotted these normalized response functions versus the
probe particle separation r scaled by the corresponding
frequency-dependent penetration depth $\delta_v = \sqrt{\eta/ \rho
\omega}$. As shown in Fig.\ \ref{fig:water normalized}, data taken at all of the different
separations $r$ ranging from $2.2~\mu m$ to $11.7~\mu m$ fall onto
a single curve for both, the parallel (Fig.\ \ref{fig:water normalized}a) and the
perpendicular (Fig.\ \ref{fig:water normalized}b) inter-particle response functions. Data
for water also collapse on water/glycerol data, after accounting
for the different viscosities, which are known in both cases.
Thus, no free fit parameters were used. The single curves of
collapsed data are in quantitative agreement with the
frequency-dependent dynamic Oseen tensor in Eqs.\ (\ref{velocity-par},\ref{velocity-perp}) shown by the solid lines in Figs.\ \ref{fig:water normalized}a and b, where the region of negative
response in the perpendicular direction corresponds again to the
back-flow region of the vortex already described above.

\begin{figure}
\centering
\includegraphics[width=8cm]{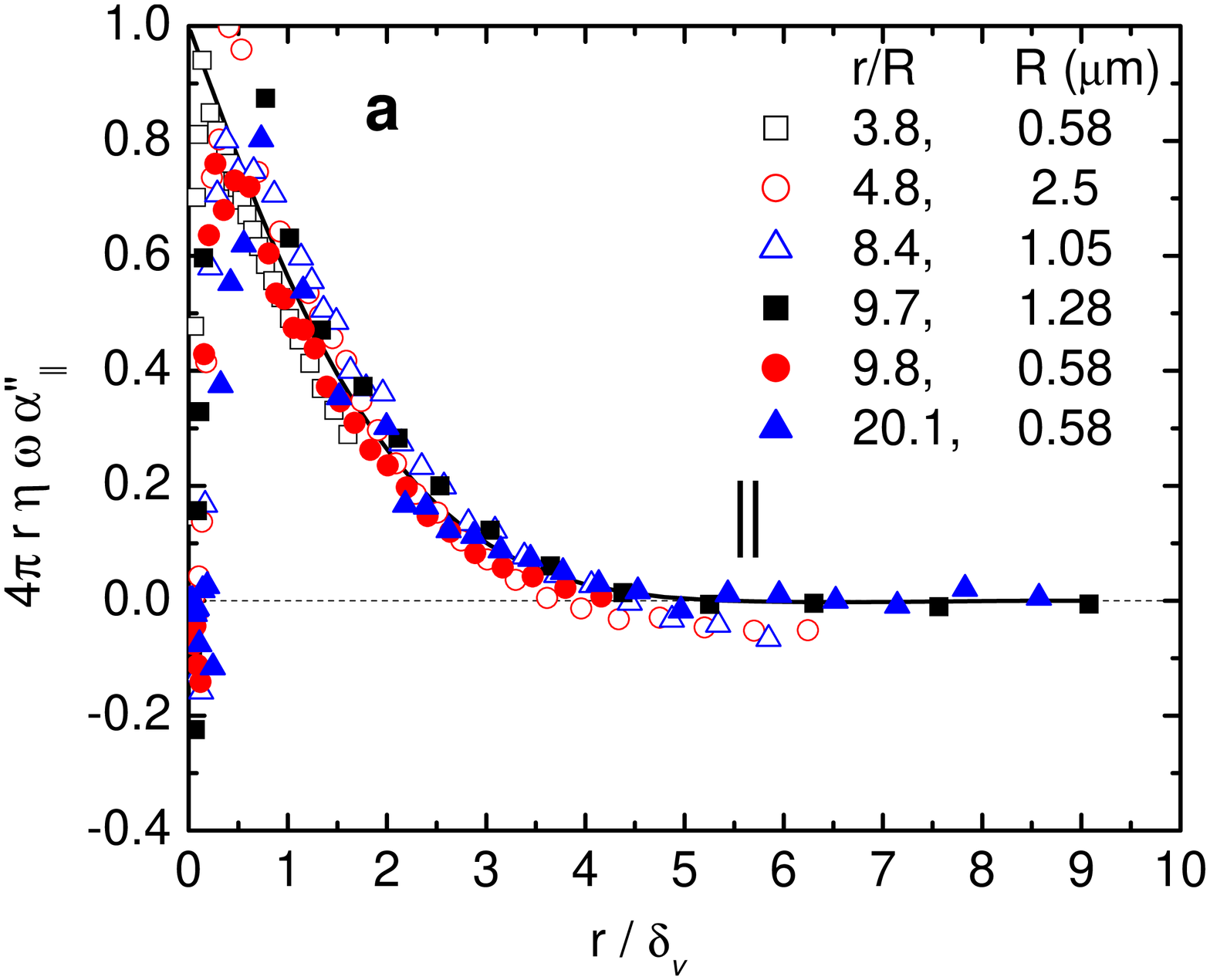}
\includegraphics[width=8cm]{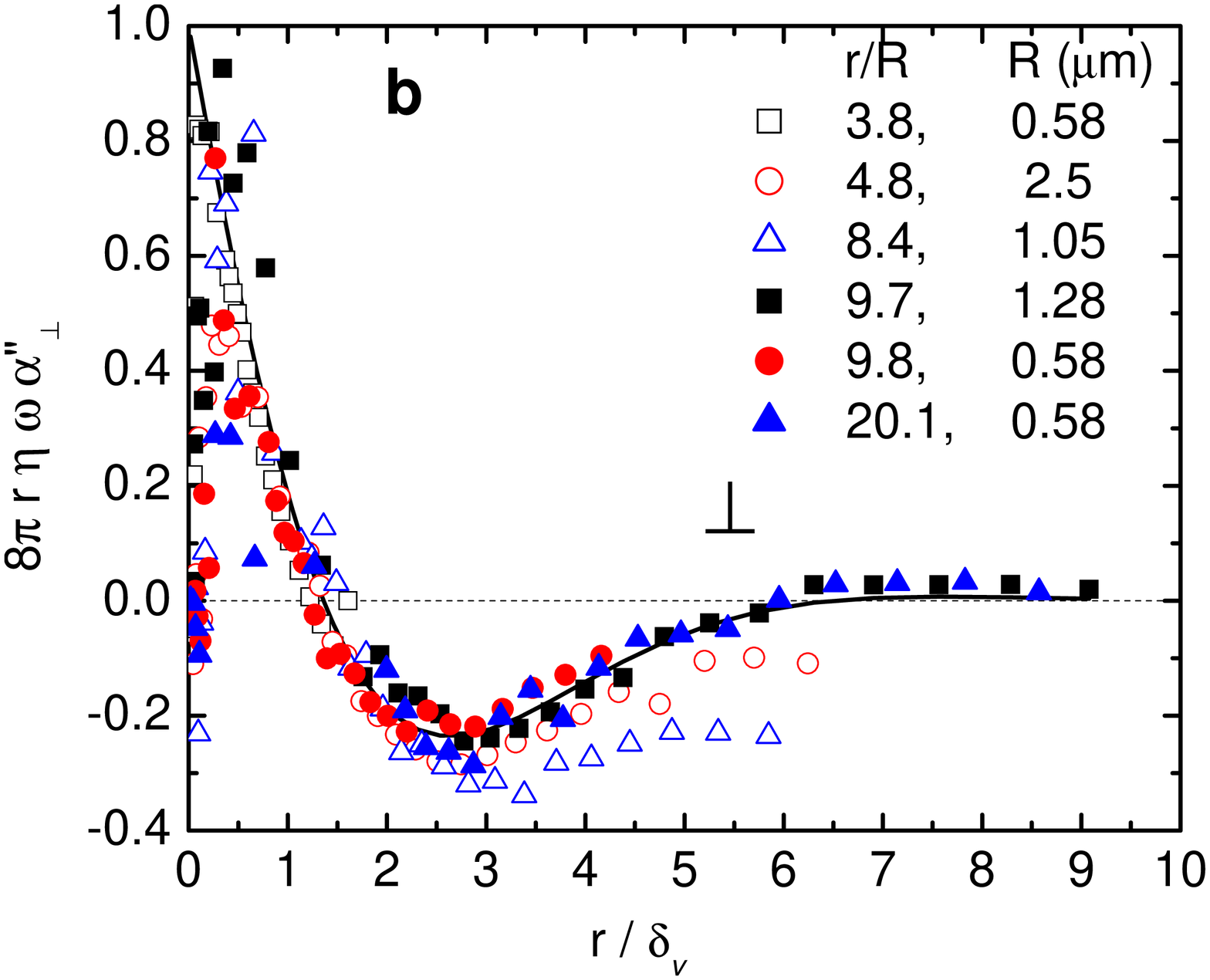}
\caption{(Color online) Normalized imaginary parts of
inter-particle displacement response functions between two probe
particles of various radii measured with the passive method in
water (silica beads, $R$ given in legend),(a) $4 \pi r \eta \omega
\alpha''_{||}$ in the parallel direction and (b) $8 \pi r \eta
\omega \alpha''_{\bot}$ in the perpendicular direction, plotted
versus the ratio of the separation distance $r$ (fixed for a given
bead pair) to the frequency-dependent viscous penetration depth
$\delta_{v}$. These response functions also represent the in-phase
velocity response normalized by the corresponding components of
the Oseen tensor. Different particle sizes (R = 0.58, 1.05, 1.28
and 2.5~$\mu$m) were used at various separation distances $r$.}
\label{fig:water R dependence}
\end{figure}

At separations $r$ large compared to particle size, the
inter-particle response functions become independent of probe
particle size and shape \cite{2fluid}, leaving a dependence only
on $r$ and $\omega$. In our analysis, we have assumed that the
particles are point-like since the ratio $r/R \geq 4$ in all
experiments. In order to directly check the validity of the
approximation, we measured the inter-particle response functions
with particles of different sizes. In Fig.\ \ref{fig:water R dependence}a and b the imaginary
parts of the normalized response functions in water $4 \pi \eta r
\omega \alpha''_{||}(\omega)$ and $8 \pi \eta r \omega
\alpha''_{\bot}(\omega)$ are plotted versus $r/\delta_v$, obtained
with probe particles of radius $R$ = 0.58$~\mu m$, 1.05~$\mu m$,
1.28~$\mu m$ and 2.5~$\mu m$. We find no systematic bead-size
dependence, justifying the point-probe approximation. Again all
normalized and scaled data collapse onto one single curve for each
channel, which in turn agrees well with the dynamic Oseen tensor.
The deviations observed in the last data points for $R$ =
1.05~$\mu m$ and 2.5~$\mu m$ are most probably due to the
influence of shot noise at high frequencies. Shot noise becomes a
problem for larger beads because of their smaller fluctuation
amplitudes, which will eventually produce signals that approach
the fixed shot noise level. The deviations observed for these
particle sizes are not consistent with a possible error due to
finite particle size, which should be larger for smaller $r/R$.

So far, we have considered the passive fluctuations that
directly measure only the imaginary (out-of phase) part of the
inter-particle response functions ($\alpha''_{||,\bot}(\omega)$).
With the active method described in Sec.\ \ref{sec:Meth}
we can determine both real
and imaginary part of the response functions. In Figs.\ \ref{fig:water Active}a and b,
we show the normalized inter-particle response functions
$\alpha'_{||,\bot}(\omega)$ and $\alpha''_{||,\bot}(\omega)$
measured by active microrheology (bead radius $R$= 0.58~$\mu m$)
in water for both parallel and perpendicular direction. In both
cases we find good agreement with Eqs.\ (\ref{velocity-par},\ref{velocity-perp}). The
slightly different separation distances in parallel ($r$ =
10.3~$\mu m$) and perpendicular ($r$ = 11.3~$\mu m$) were due to
different settings of the AOD signal in these measurements.

\subsection{Viscoelastic solutions}

\begin{figure}
\centering
\includegraphics[width=8cm]{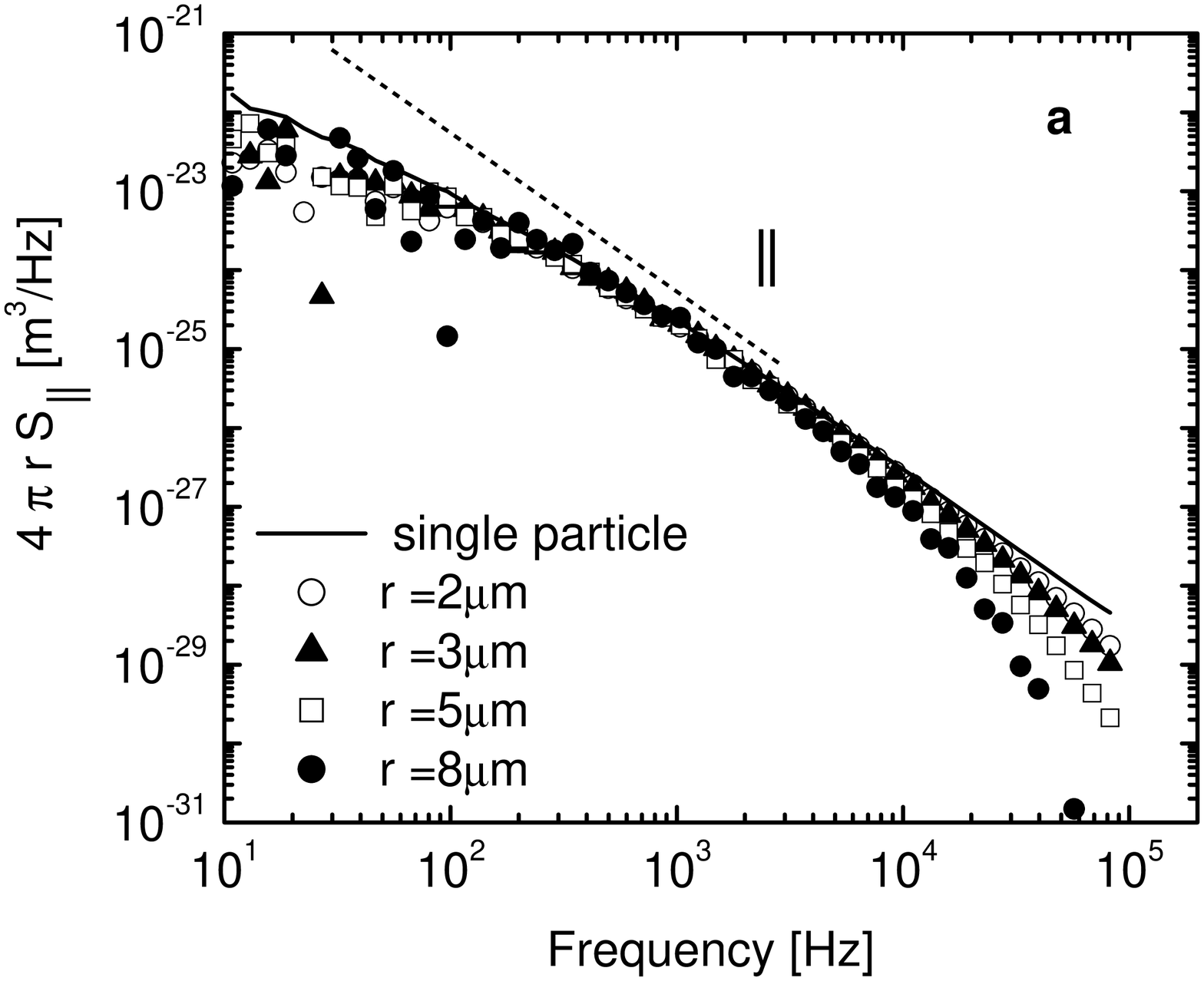}
\includegraphics[width=8cm]{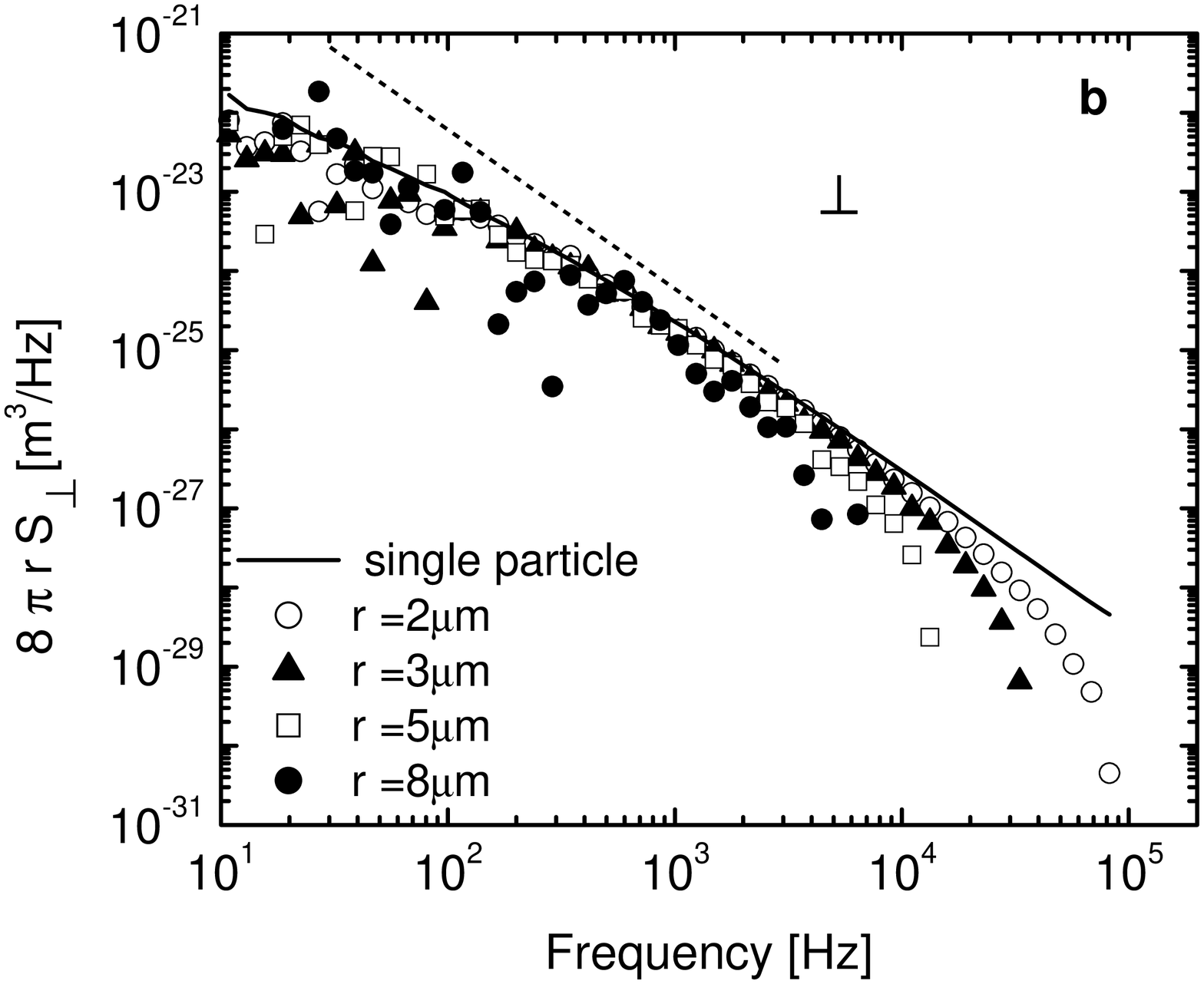}
\caption{Normalized displacement cross-correlation functions $4
\pi r S_{||}$ (a) and $8 \pi r S_{\bot}$ (b) of two probe
particles (silica beads, $R$ = 0.58 $\mu$m) in worm-like micelle
solutions (c$_{m}$ = 1~wt\%) versus frequency ($f = \omega/2\pi$)
compared for different separation distances $r$. The solid lines represent
the auto-correlation function of a single particle normalized
by $6 \pi R$. The dashed lines indicate slopes of -2, corresponding to diffusive 
motion.} \label{fig:micelles spectra}
\end{figure}

In viscoelastic polymer solutions, the elastic component in the
response of the medium modifies the propagation of the inertial
vortex. We first discuss our results for worm-like micelle
solutions, the viscoelastic properties of which have been
characterized by microrheology \cite{micelles, Atakhorrami
micelles}. In Fig.\ \ref{fig:micelles spectra} we have plotted the displacement
cross-correlation functions ($4 \pi r S_{||}$ and $8 \pi r
S_{\bot}$) in a 1\% worm-like micelle solution versus frequency for
two particles at various separation distances $r$ between 2 and
$8~\mu m$. For comparison, we have added the scaled
auto-correlation function $6 \pi R S_{auto}$ for a single particle
($R$ = 0.58~$\mu m$). In contrast to the situation in simple
liquids, the particles were here confined by the surrounding
polymer network and do not diffuse freely. Thus, the frequency
dependence is weaker than for Brownian motion (i.e., the slope is
less steep than -2 in the low-frequency, non-inertial regime) 
\cite{micelles, Atakhorrami micelles}. As
before, the displacement cross-correlation functions of the two
probe particles are used to map the vortex-like flow pattern and
its propagation in time. The $r$-dependent decrease of the
cross-correlation functions occurs for both parallel (Fig.\ \ref{fig:micelles spectra}a) and
perpendicular (Fig.\ \ref{fig:micelles spectra}b) directions, although the latter is more
apparent.

\begin{figure}
\centering
\includegraphics[width=8cm]{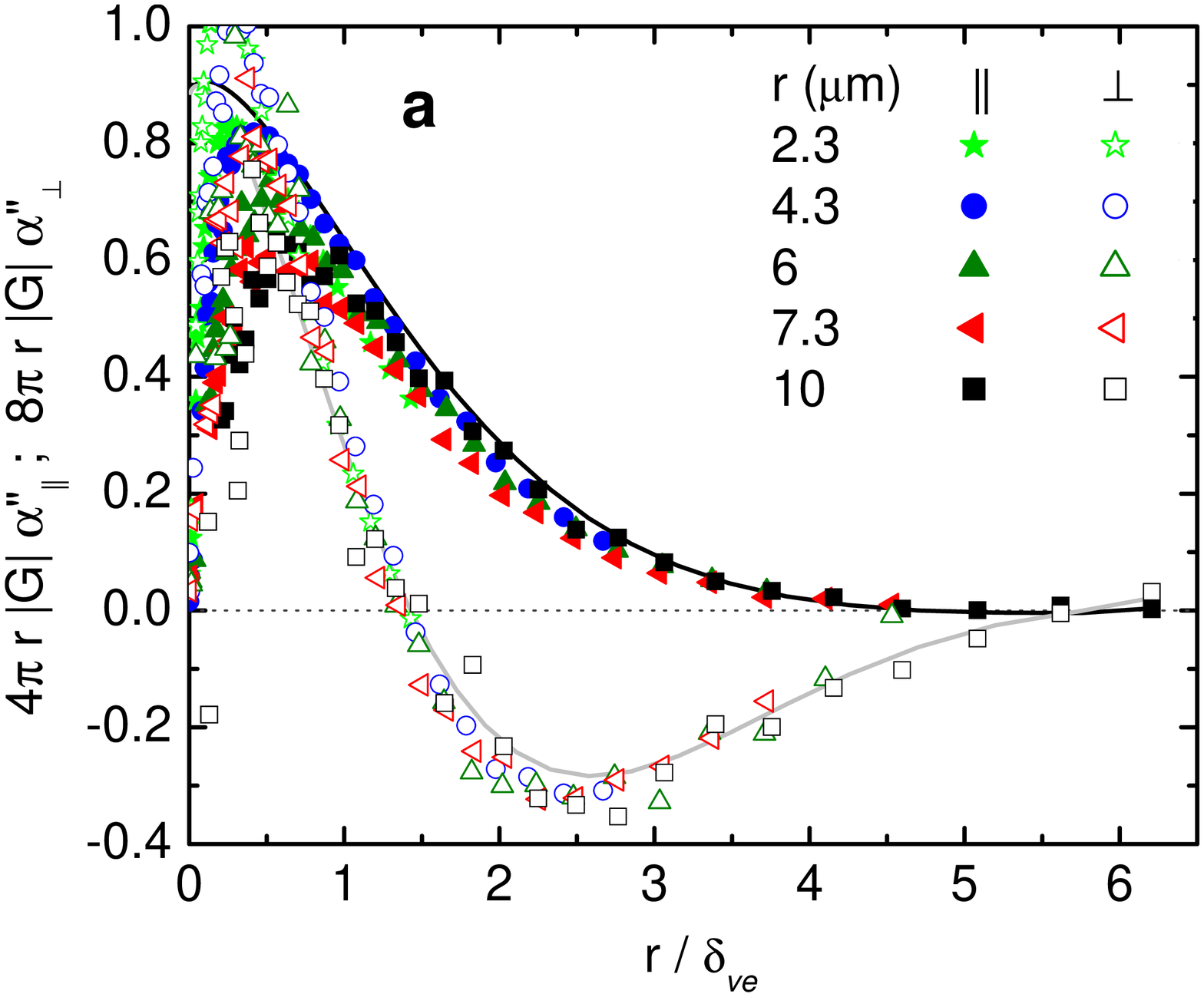}
\includegraphics[width=8cm]{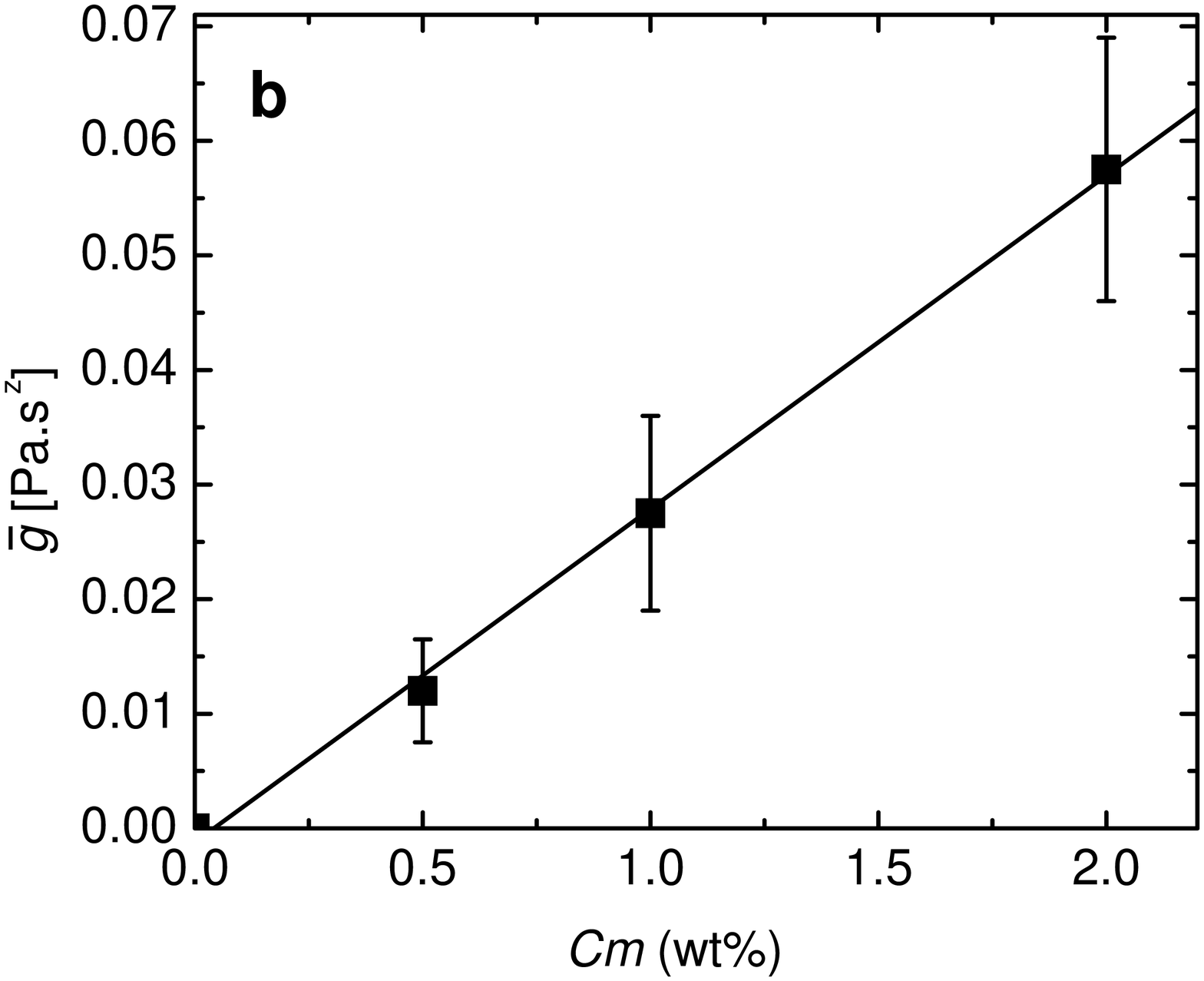}
\caption{(Color online) (a) Collapse of the imaginary parts of the
normalized inter-particle response functions ($4 \pi r |G|
\alpha''_{||}(\omega)$ and $8 \pi r |G| \alpha''_{\bot}(\omega)$)
between two probe particles (silica beads, $R$ = 0.58 $\mu$m),
measured with the passive method for different separation
distances $r$ in worm-like micelle solutions of 1~wt\%, plotted
versus the ratio of the separation distance $r$ (fixed for a given
bead pair) to the frequency-dependent viscoelastic penetration
depth $\delta_{ve}$. Here, the viscoelastic penetration depths
were determined by varying the parameters $\bar g$ and $z$ in Eqs.\
(\ref{alpha-par}-\ref{chi_perp},\ref{Gomega}) to obtain collapse. Optimal parameters were $z$ = 0.68
$\pm$ 0.05 and $\bar g$ = 0.0275 $\pm$ 0.008~Pa s$^{z}$, where the
solvent (water) viscosity has been taken into account. (b)
Dependence of (optimal) parameter $\bar g$ on micelle concentration for
a fixed $z$.} \label{fig:micelle collapse}
\end{figure}

In order to collapse these data onto a single curve for each of
the two channels (parallel and perpendicular), following Eqs.\
(\ref{alpha-par}-\ref{chi_perp}), we plot the normalized inter-particle response functions $4
\pi |G| r \alpha''_{||}$ and $8 \pi |G| r \alpha''_{\bot}$ versus
scaled $r/\delta_{ve}$ , where $\delta_{ve}$ is the viscoelastic
penetration depth. Unlike for the water and water/glycerol
samples, we do not \emph{a priori} know the frequency-dependent
shear modulus $G^\star(\omega)$. Based on theoretical expectations
for flexible polymers \cite{Doi}, as well as on prior
high-frequency rheology of worm-like micelle solutions
\cite{micelles, Atakhorrami micelles}, we assume that the shear
modulus has the functional form given in Eq.\ (\ref{Gomega}). Thus, in order
to achieve the collapse of all data onto the master curves
represented by Eqs.\ (\ref{alpha-par}-\ref{chi_perp}), we vary the two correlated
parameters $\bar g$ and $z$. We expect $z$ to be independent of the
micelle concentration, while $\bar g$ should depend linearly on the
polymer/micelle concentration. In Fig.\ \ref{fig:micelle collapse}a, we show the resulting
collapse of the normalized inter-particle response functions in
parallel and perpendicular directions for different separations
$r$. The predictions of Eqs.\ (\ref{alpha-par}-\ref{chi_perp}) are shown with black and
gray lines \cite{liverpool}. The best overall collapse of the data
for worm-like micelles solutions at all concentrations (0.5, 1 and
2 weight percent) and separations $r$ from $2~\mu m$ to $16~\mu m$
was found for $z$ = 0.68 $\pm$ 0.05 and $\bar g$ concentration
dependence as seen in Fig.\ \ref{fig:micelle collapse}b.

\begin{figure}
\centering
\includegraphics[width=8cm]{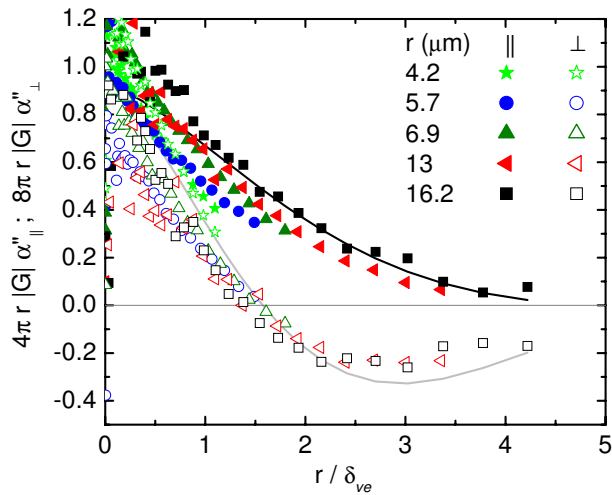}
\caption{(Color online) Collapse of the imaginary parts of the
normalized inter-particle response functions between two probe
particles (silica beads, $R$ = 0.58 $\mu$m), measured with the
passive method for different separation distances $r$ in F-actin
solutions of concentration 1 mg/ml, plotted versus the ratio of
the separation distance $r$ (fixed for a given bead pair) to the
frequency-dependent viscoelastic penetration depth $\delta_{ve}$.
Here, the parameters $\bar g$ and $z$ in Eq.\ (\ref{Gomega}) were varied to
obtain simultaneous collapse of all data sets onto Eqs.\ (\ref{alpha-par}-\ref{chi_perp}),
using a single set of parameters $z$ and $\bar g$, while accounting for
the solvent (water) viscosity. We find $z$ = 0.78 $\pm$ 0.1, and
$\bar g$ = 0.18 $\pm$ 0.13 as optimal parameters. Data are presented
for parallel (closed symbols) and perpendicular (open symbols)
directions. The solid black line represents Eq.\ (\ref{im_alpha_parr}) and the
gray line represents Eq.\ (\ref{im_alpha_perp}), both with $z$ = 0.75 and $\bar g$ =
0.2~Pa s$^{z}$. } \label{fig:Actin collapse}
\end{figure}

In order to further test the inertial effects in viscoelastic
media, we also performed experiments on another viscoelastic fluid
with a somewhat different frequency dependent shear modulus,
namely entangled F-actin solutions. At high frequency, semiflexible
F-actin filaments contribute to the viscoelasticity of the medium
in a different way from flexible polymers \cite{Morse98, Gittes98,
MR97, Koenderink actin}. Therefore, the spatial structure and
the propagation dynamics of the vortex should be different. Figure
\ref{fig:Actin collapse} shows the collapse of the inter-particle response functions $4
\pi |G| r \alpha''_{||}$ and $8 \pi |G| r \alpha''_{\bot}$ plotted
versus scaled distance $r/\delta_{ve}$ onto two master curves for
the parallel and the perpendicular direction. The actin
concentration was 1mg/ml and the probe radius 0.58~$\mu m$ and we
used separation distances $r$ ranging from 4.2~$\mu m$ to
16.2~$\mu m$. In an F-actin solution of this concentration, the
magnitude of the shear modulus is large, therefore the vortex
propagates faster, making it harder to observe. In particular, it
was difficult to determine the parameters $\bar g$ and $z$ in this
case. We found the best collapse with $z$ = 0.78 $\pm$ 0.1 for
data taken in passive method. We then fixed $z$ to 0.75 known from
the power law dependence behavior reported previously
\cite{Morse98, Gittes98, MR97, Koenderink actin}, and found
$\bar g$ = 0.18 $\pm$ 0.13Pa s$^{z}$. To reduce the large
error bars in the passive method, it would be necessary to repeat
our measurements at higher frequencies and/or larger separation
distances. Nevertheless, our results are consistent with prior
measurements and predictions of both parameters.

Independently, we have measured both real,
$\alpha'_{||,\bot}(\omega)$, and imaginary,
$\alpha''_{||,\bot}(\omega)$, parts of the response functions
directly by actively manipulating one particle and measuring the
response of the other. In Figs.\ \ref{fig:actin Active}a and b, the parallel (Fig.\ \ref{fig:actin Active}a)
and perpendicular (Fig.\ \ref{fig:actin Active}b) complex inter-particle response
functions for $c$ = 1~mg/ml F-actin solutions are shown, probed
with beads of 1.28~$\mu m$ radius. Here the inter-particle
response functions were fitted with Eqs.\ (\ref{alpha-par}-\ref{chi_perp}) to find
parameters $\bar g$ = 0.22$\pm$0.05 Pa s$^{z}$ and $z$= 0.78$\pm$0.01
simultaneously. Data for both parallel ($r$ = 12.1~$\mu m$) and
perpendicular ($r$ = 13.5~$\mu m$) channels are compared with $z$
= 0.75 and $\bar g$ = 0.22 Pa s$^{z}$ in Fig.\ \ref{fig:actin Active}a and b.

The values of $z$ and $\bar g$ found from both methods are consistent
and also agree with results from prior experimental microrhelogy
and macrorheology experiments for both entangled actin solutions
\cite{MR97, Morse98, Gittes98, Gardel, Koenderink actin} and
worm-like-micelle solutions \cite{micelles, Atakhorrami micelles}.
To obtain these values it was essential to model the inertial
effects including both polymer and solvent contributions to the
shear modulus. We observed that, although the high-frequency
rheology of the polymer solution is dominated by the polymer, the
background solvent contributes non-negligibly to the inertial
vortex propagation. To test this, we excluded the solvent shear
modulus ($-i \omega \eta$) in Eqs.\ (\ref{alpha-par}-\ref{chi_perp},\ref{Gomega}), and analyzed our data
assuming a high-frequency shear modulus of the form $G = \bar g \omega
^{z}$. For both worm-like micelle solutions and entangled F-actin
solutions, we found much larger values of $z \sim$ 0.9 and a
nonlinear concentration dependence of $\bar g$, contrary to
expectations.

\section{DISCUSSION}

In our experiments we have directly resolved the inertial
response/flow of fluids on micrometer and microsecond time scales
using optical trapping and interferometric particle-tracking. Our
results demonstrate that vorticity and stress propagate
diffusively in simple liquids and super-diffusively in
viscoelastic media. One consequence of inertial vortex formation
is the long-time tail effect observed in light scattering
experiments \cite{exper_tails}. To connect to these results, we
calculated the velocity auto-correlation of a single particle from
the displacement fluctuations. Unfortunately, the effect we are
looking for is subtle and is difficult to detect in the presence
of other factors. At the highest frequencies the vortex is still
influenced by the finite probe size, and at intermediate
frequencies the particle motion is already affected by the laser
trap potential. The results were thus inconclusive. Similar
problems have been reported in Ref. \cite{Lukic}. The effects of
inertia we have described here set a fundamental limit to the
applicability of two-particle microrheology techniques which are
based on the measurement of cross-correlated position fluctuations
of particles \cite{crocker, Gardel, 2fluid, micelles,Koenderink
actin}. Inertia limits the range of stress propagation at high
frequencies, stronger in soft media such as those studied here
than in media with higher viscoelastic moduli. Inertia affects
measurements at frequencies as low as 1~kHz for separations of
order 10 micrometers, showing an apparent increase of the measured
shear moduli below their actual values \cite{Atakhorrami
micelles}. Since the stress propagation is diffusive, or nearly
so, even measurements at video rates can be affected for probe
particle separations of order 50 micrometers. As we have shown
here, these inertial effects are more pronounced in the
perpendicular inter-particle response functions than in the
parallel ones. This suggests that one should obtain shear moduli
from the parallel inter-particle response functions if one doesn't
want to correct for inertia. In a more precise analysis of
two-particle microrheology experiments, the fluid response
function can not simply be modeled by a generalized
Stokes-Einstein relationship and has to be corrected for inertial
effects according to the probed frequency as well as the particles
separation. Such corrections, however, will necessarily be
limited, given the exponential attenuation of stress due to
inertia.

\section{Acknowledgments}

Actin was purified by K.C. Vermeulen and silica particles were
kindly donated by C. van Kats (Utrecht University). F. Gittes, J.
Kwiecinska, and J. van Mameren helped with developing the data
analysis software. We thank D. Frenkel, W. van Saarloos, A.J.
Levine and K.M. Addas for helpful discussions. This work was
supported by the Foundation for Fundamental Research on Matter
(FOM). Further support (C.F.S.) came from the DFG Center for
the Molecular Physiology of the Brain (CMPB) and the
DFG Sonderforschungsbereich 755.


\begin{thebibliography}{99}


\bibitem{LandauFluids} L.D.\ Landau and E.M.\ Lifshitz, {\em Fluid Mechanics}, Butterworth-Heinemann (Oxford, 2000).

\bibitem{brenner} J.\ Happel and H.\ Brenner, {\em Low Reynolds Number Hydrodynamics}, McGraw-Hill  (New York, 1963)

\bibitem{Guyon} E.\ Guyon et al, {\em Physical Hydrodynamics}, Oxford
  University Press (New York, 2001)

\bibitem{Oseen} C.W.\ Oseen, Hydrodynamik (Akademische Verlagsgesellschaft, Leipzig, 1927), p.\ 47.

\bibitem{Alder} B.J.\ Alder and T.E.\ Wainwright, Phys.\ Rev.\ A {\bf 1}, 18 (1970).

\bibitem{exper_tails} J.P.\ Boon and A.\ Bouiller,
Phys.\ Lett.\ {\bf 55A}, 391 (1967); G.L.\ Paul and P.N.\ Pusey,
J.\ Phys.\ A {\bf 14}, 3301 (1981); K.\ Ohbayashi, T.\ Kohno, and
H.\ Utiyama, Phys.\ Rev.\ A\ {\bf 27}, 2632 (1983); D.A.\ Weitz,
D.J.\ Pine, P.N.\ Pusey, and R.J.A.\ Tough, Phys.\ Rev.\ Lett.\
{\bf 63}, 1747 (1989).

\bibitem{maryam} M.\ Atakhorrami, G.H.\ Koenderink, C.F.\ Schmidt, and F.C.\
MacKintosh, Phys.\ Rev.\ Lett.\ {\bf 95}, 208302 (2005).

\bibitem{theor_tails} R.\ Zwanzig and M.\ Bixon, Phys.\ Rev.\ A \textbf{2}, 2005 (1970);
M.H.\ Ernst, E.H.\ Hauge, and J.M.J.\ van Leeuwen, Phys.\ Rev.\ Lett.\ \textbf{25}, 1254 (1970);
J.R.\ Dorfman and E.G.D.\ Cohen, Phys.\ Rev.\ Lett.\ \textbf{25}, 1257 (1970);
D.\ Bedeaux and P.\ Mazur, Phys.\ Lett.\ {\bf 43A}, 401 (1973);
D.\ Bedeaux and P.\ Mazur, Physica (Amsterdam) \textbf{73}, 431 (1974).

\bibitem{Lukic} B.\ Lukic, S.\ Jeney, C.\ Tischer, et al.,
Phys.\ Rev.\ Lett.\ {\bf 95}, 160601 (2005).

\bibitem{Morkel} C.\ Morkel, C.\ Gronemeyer, W.\ Glaser, J.\ Bosse, Phys.\
Rev.\ Lett.\ {\bf 58}, 1873 (1987).

\bibitem{liverpool} T.B.\ Liverpool and F.C.\ MacKintosh, Phys.\ Rev.\ Lett.\ {\bf 95},
208303 (2005).

\bibitem{detection Gittes} F.\ Gittes and C.F.\ Schmidt, Optics
Lett.\ {\bf 23}, 7 (1998).

\bibitem{setup maryam} M.\ Atakhorrami, K.M. Addas and C.F.\ Schmidt,
unpublished.

\bibitem{crocker} J.C.\ Crocker, M.T.\ Valentine, E.R.\ Weeks, T.\ Gisler, P.D.\ Kaplan, A.G.\ Yodh, and D.A.\ Weitz, Phys.\ Rev.\ Lett.\ {\bf 85}, 888 (2000).

\bibitem{meiners} J-C.\ Meiners and S.R.\ Quake, Phys.\ Rev.\ Lett.\ {\bf 82}, 2211 (1999).

\bibitem{bartlett} S.\ Henderson, S.\ Mitchell, and P.\ Bartlett, 
Phys.\ Rev.\ E {\bf 64}, 061403 (2001)

\bibitem{Starrs} L.\ Starrs and P.\ Bartlett, Journal of Physics-Cond.\ Mat.\ {\bf 15}, S251 (2003).

\bibitem{Gardel} M.L.\ Gardel, M.T.\ Valentine, J.C.\ Crocker, A.R.\ Bausch and D.A.\ Weitz, 
Phys.\ Rev.\ Lett.\ {\bf 91}, 158302 (2003).

\bibitem{Koenderink actin} G.H.\ Koenderink, M.\ Atakhorrami, F.C.\ MacKintosh, C.F.\ Schmidt, Phys.\ Rev.\ Lett.\ \textbf{96}:138307 (2006).

\bibitem{MR97} F.\ Gittes, B.\ Schnurr, P.D.\ Olmsted, F.C.\ MacKintosh and C.F.\ Schmidt, 
Phys.\ Rev.\ Lett.\ {\bf 79}, 3286 (1997); B.\ Schnurr, F.\ Gittes, F.C.\ MacKintosh, C.F.\ Schmidt, Macromolecules {\bf 30}, 7781 (1997).

\bibitem{Mizuno active} D.\ Mizuno, C.\ Tardin, C.F.\ Schmidt, F.C.\ MacKintosh, Science \textbf{315}:370 (2007).

\bibitem{Hough} L.A.\ Hough and H.D.\ Ou-Yang, Phys.\ Rev.\ E {\bf 65}, 021906 (2002).

\bibitem{bird} R.B.\ Bird, R.C.\ Armstrong, O.\ Hassager, {\em Dynamics of Polymeric Liquids}, Wiley (New york, 1987).

\bibitem{2fluid} A.J.\ Levine and T.C.\ Lubensky, Phys.\ Rev.\ Lett.\ {\bf 85}, 1774 (2000); A.J.\ Levine and T.C.\ Lubensky, Phys.\ Rev.\ E {\bf 63}, 041510 (2001).

\bibitem{Brochard} F.\ Brochard and P.G.\ de Gennes, Macromolecules \textbf{10}, 1157 (1977); S.T.\ Milner, Phys.\ Rev.\ E \textbf{48}, 3674 (1993).

\bibitem{Mazur} D.\ Bedeaux and P.\ Mazur, Physica (Amsterdam) {\bf 76}, 235 (1974); {\bf 76}, 247 (1974).

\bibitem{Doi} M.\ Doi and S.F.\ Edwards, {\it The theory of polymer dynamics}, Oxford University Press (New York, 1986).

\bibitem{Morse98} D.C.\ Morse, Macromolecules {\bf 31}, 7030 (1998); {\bf 31}, 7044 (1998).

\bibitem{Gittes98} F.\ Gittes and F.C.\ MacKintosh, Phys.\ Rev.\ E {\bf 58}, R1241 (1998).

\bibitem{NoteWater} For the case of polymer solutions, this represents only the polymer contribution to the shear modulus. For comparison with experiment \cite{maryam} we have also taken into account the solvent contribution in $G(\omega)=\bar{g}(-i\omega)^z-i\omega\eta$.

\bibitem{podlubny} I.\ Podlubny, {\it Fractional Differential
    equations}, Academic Press (London, 1999).

\bibitem{erdelyi} A.\ Erd\'elyi (ed.), {\it Higher  Transcendental Functions, vol.\ 3}, McGraw-Hill (New York, 1953).

\bibitem{paris} R.B.\ Paris, Proc.\ Roy.\ Soc.\ A., {\bf 458}, 3041 (2002).

\bibitem{Berret} J.F.\ Berret, J.\ Appell, and G.\ Porte, Langmuir{\bf 9}, 2851 (1993).

\bibitem{Pardee} J.D.\ Pardee and J.A.\ Spudich, {\em in Structural and Contractile Proteins} (PartB: The Contractile Apparatus and the Cytoskeleton), ed.\ by D.W.\ Frederiksen and L.W.\ Cunningham (Academic Press, Inc., San Diego, 1982), Vol.\ 85, p. 164.

\bibitem{Allersma} M.W.\ Allersma, F.\ Gittes, M.J.\ deCastro, et al., Biophys.\ J.\ {\bf 74}, 1074 (1998).

\bibitem{MizunoTechnical} D.\ Mizuno, D.A. Head, F.C.\ MacKintosh and C.F.\ Schmidt,
unpublished.

\bibitem{Erwin detection} E.J.G.\ Peterman, M.A.\ van Dijk, L.C.\ Kapitein, et al., Rev.\ Scien.\ Ins.\ {\bf 74}: 3246 (2003).

\bibitem{signal and noise} F.\ Gittes and C.F.\ Schmidt, in Methods in Cell Biology (Academic Press, 1998), Vol. 55, p. 129.

\bibitem{Landau stat} L.D.\ Landau and E.M.\ Lifshitz and L.P.\ Pitaevskii {\em Statistical Physics}, (Pergamon Press, Oxford, New York, 1980).

\bibitem{Mason} T.G.\ Mason and D.A.\ Weitz, Phys.\ Rev.\ Lett.\ {\bf 75}: 2770 (1995).

\bibitem{micelles} M.\ Buchanan, M.\ Atakhorrami, J.F.\ Palierne, F.C.\ MacKintosh and C.F.\ Schmidt, 
Phys.\ Rev.\ E {\bf 72}, 011504 (2005); 
M.\ Buchanan, M.\ Atakhorrami, J.F.\ Palierne, C.F.\ Schmidt, Macromolecules {\bf 38}, 8840 (2005).

\bibitem{Atakhorrami micelles} M.\ Atakhorrami and C.F.\ Schmidt,
Rheol. Acta, {\bf 45}, 449 (2006).

\end{thebibliography}
\end{document}